\documentclass[12pt]{article}

\usepackage{amsmath}
\usepackage{graphicx}
\usepackage{enumerate}
\usepackage{natbib}
\usepackage{url} 
\usepackage{nicematrix}
\usepackage{multibib}
\usepackage{hyperref}
\usepackage{centernot}
\usepackage{doi}
\usepackage{subcaption}
\usepackage{booktabs}

\usepackage{tocloft}
\usepackage{etoolbox}
\usepackage{titletoc}
\usepackage{tabularx}
 \usepackage{booktabs}
 \usepackage{multirow}
\usepackage{siunitx}

\newcites{supp}{Supplementary Material References}
\newcommand{\blind}{1}
\def\spacingset#1{\renewcommand{\baselinestretch}%
	{#1}\small\normalsize} \spacingset{1}

\usepackage{amsthm}
\newtheoremstyle{myexamplestyle} 
  {0pt}                         
  {0pt}                         
  {\normalfont}                 
  {}                           
  {\bfseries}                  
  {.}                          
  { }                          
  {\thmname{#1} \thmnumber{#2}: \textbf{#3}} 

\theoremstyle{myexamplestyle}
\newtheorem{example}{Example}

\usepackage[mathscr]{eucal}

\usepackage{dlfltxbcodetips,bbm,mathtools,amsfonts,amsmath,amssymb,fancybox,graphicx,bm,latexsym}

\usepackage{tikz}
\usetikzlibrary{bayesnet}
\usetikzlibrary{fit,positioning}

\newcommand{\bY}{\mathbf{Y}}
\newcommand{\bJ}{\mathbf{J}}
\newcommand{\bG}{\mathbf{G}}

\newcommand{\by}{\mathbf{y}}

\usepackage[tikz]{bclogo}

\newcommand{\bt}{
 \begin{bclogo}[logo= \bcloupe 
,couleur={rgb:orange,0;yellow,0;white,1},arrondi=0.1,ombre=false]}
	\newcommand{\et}{\end{bclogo}\s}
\newcommand{\btt}{\begin{box}}
	\newcommand{\ett}{\end{box}}

\newcommand{\btheorem}{\begin{bclogo}[couleur={rgb:orange,0;yellow,0;white,1},arrondi=0.1,logo=\bcplume,ombre=true]{Theorem}}
	\newcommand{\ettheorem}{\end{bclogo}}

\newcommand{\bsh}{\begin{bclogo}[couleur={rgb:orange,0;yellow,0;white,1},arrondi=0.1,logo=\bcpanchant,ombre=true]}
	\newcommand{\esh}{\end{bclogo}}

\DeclareMathOperator*{\argmax}{arg\,max}

\newcommand{\benum}{\begin{enumerate}}
	\newcommand{\eenum}{\end{enumerate}}

\newcommand{\bq}{\begin{quote}\em}
	\newcommand{\eq}{\end{quote}}
\newcommand{\bbq}{\begin{quote}\bf\em}
	\newcommand{\ebq}{\end{quote}}

\renewcommand{\bar}{\overline}

\newcommand{\mbR}{\mathbb{R}}

\newcommand{\mbE}{\mathbb{E}}

\newcommand{\mbP}{\mathbb{P}}

\newcommand{\hide}[1]{}
\newcommand{\ghost}[1]{}
\newcommand{\ba}{\begin{array}{llllllllll}}
	\newcommand{\ea}{\end{array}}
\newcommand{\bea}{\begin{equation}\begin{array}{llllllllll}}
		\newcommand{\eea}{\end{array}\end{equation}}

\newcommand{\beno}{\begin{equation}\begin{array}{llllllllll}\nonumber}
		\newcommand{\be}{\begin{equation}\begin{array}{llllllllll}}
				\newcommand{\ee}{\end{array}\end{equation}}
		\newcommand{\bi}{\begin{itemize}}
			\newcommand{\ei}{\end{itemize}}
		\newcommand{\ben}{\begin{enumerate}}
			\newcommand{\een}{\end{enumerate}}

		\newcommand{\dsum}{\displaystyle\sum\limits}
		
		\newcommand{\dprod}{\displaystyle\prod\limits}

		\newcommand{\s}{\vspace{0.25cm}}
		
		\newcommand{\bx}{\bm{x}}

		\newcommand{\bA}{\bm{A}}
		
		\newcommand{\bv}{\bm{v}}

		\newcommand{\mI}{\mathbb{I}}

		\newcommand{\bu}{\bm{u}}

		\newcommand{\bs}{\bm{s}}
        \newcommand{\bh}{\bm{h}}
		\newcommand{\bz}{\bm{z}}
		\newcommand{\bZ}{\bm{Z}}
		
		\newcommand{\balpha}{\bm{\alpha}}
		
		\newcommand{\bbeta}{\bm{\beta}}
		\newcommand{\bgamma}{\mbox{\boldmath$\gamma$}}

		\newcommand{\bSigma}{\mbox{\boldmath$\Sigma$}}
		\newcommand{\bTheta}{\bm\Theta}
		
		\newcommand{\btheta}{\boldsymbol{\theta}}

		\newcommand{\bV}{\bm{V}}

		\newcounter{counterexample}

		\newcounter{definition}
		
		\newcounter{theorem}

		\newcounter{proposition}

		\newcounter{result}

		\newcounter{tproof}
		\newcounter{corollary}

		\newcounter{cproof}
		\newcounter{lemma}

		\newcounter{com}

		\newcounter{lproof}
		\newcounter{assumption}
		\newenvironment{assumption}[1][]{\refstepcounter{assumption}\par\medskip\indent%
			\textbf{Condition~\theassumption #1}. \rmfamily}{\vspace{-3em}}

		\newif\ifmydraft
		\mydraftfalse

		\ifmydraft
		
		\pagecolor{black!20}
		\else
		
		\fi

		\ifmydraft
		
		\pagecolor{black!20}
		\fi
		
		\usepackage{multicol}

		\makeatletter
		\newcommand*{\deq}{\mathrel{\rlap{%
					\raisebox{0.3ex}{$\m@th\cdot$}}%
				\raisebox{-0.3ex}{$\m@th\cdot$}}=}
		\makeatother

\definecolor{spdred}{RGB}{227,0,15}

\newcommand{\triadppp}{
\begin{tikzpicture}[scale=0.25, baseline={([yshift=-.5ex]current bounding box.center)}]
\coordinate (A) at (90:1);
\coordinate (B) at (210:1);
\coordinate (C) at (330:1);
\draw[thick,green!70!black] (A) -- node[black,above left=-1pt] {\tiny$+$} (B);
\draw[thick,green!70!black] (B) -- node[black,below=1pt] {\tiny$+$} (C);
\draw[thick,green!70!black] (C) -- node[black,above right=-1pt] {\tiny$+$} (A);
\foreach \p in {A,B,C} {
  \fill (\p) circle (4pt);
}
\end{tikzpicture}
}

\newcommand{\triadmmm}{
\begin{tikzpicture}[scale=0.25, baseline={([yshift=-.5ex]current bounding box.center)}]
\coordinate (A) at (90:1);
\coordinate (B) at (210:1);
\coordinate (C) at (330:1);
\draw[thick,spdred] (A) -- node[black,above left=-1pt] {\tiny$-$} (B);
\draw[thick,spdred] (B) -- node[black,below=1pt] {\tiny$-$} (C);
\draw[thick,spdred] (C) -- node[black,above right=-1pt] {\tiny$-$} (A);
\foreach \p in {A,B,C} {
  \fill (\p) circle (4pt);
}
\end{tikzpicture}
}

\newcommand{\triadppm}{
\begin{tikzpicture}[scale=0.25, baseline={([yshift=-.5ex]current bounding box.center)}]
\coordinate (A) at (90:1);
\coordinate (B) at (210:1);
\coordinate (C) at (330:1);
\draw[thick,green!70!black] (A) -- node[black,above left=-1pt] {\tiny$+$} (B);
\draw[thick,green!70!black] (C) -- node[black,above right=-1pt] {\tiny$+$} (A);
\draw[thick,spdred] (B) -- node[black,below=1pt] {\tiny$-$} (C);
\foreach \p in {A,B,C} {
  \fill (\p) circle (4pt);
}
\end{tikzpicture}
}

\newcommand{\triadpmm}{
\begin{tikzpicture}[scale=0.25, baseline={([yshift=-.5ex]current bounding box.center)}]
\coordinate (A) at (90:1);
\coordinate (B) at (210:1);
\coordinate (C) at (330:1);
\draw[thick,green!70!black] (A) -- node[black,above left=-1pt] {\tiny$+$} (B);
\draw[thick,spdred] (C) -- node[black,above right=-1pt] {\tiny$-$} (A);
\draw[thick,spdred] (B) -- node[black,below=1pt] {\tiny$-$} (C);
\foreach \p in {A,B,C} {
  \fill (\p) circle (4pt);
}
\end{tikzpicture}
}

\begin{document}

\def\spacingset#1{\renewcommand{\baselinestretch} {#1}\small\normalsize} \spacingset{1}
\date{}

\spacingset{1.5}

\if1\blind
{
\title{\bf Scalable Signed Exponential Random Graph Models under Local Dependence}
\author{Marc Schalberger\thanks{Corresponding author: m.schalberger@fu-berlin.de. Postal address: Freie Universität Berlin, Garystr. 21, Berlin 14195, Germany}, Freie Universität Berlin \\
Cornelius Fritz, Trinity College Dublin \\}
\maketitle
} \fi

\if0\blind
{
\title{\bf Scalable Signed Exponential Random Graph Models under Local Dependence}
\maketitle
} \fi

\vspace{-1cm}

\spacingset{1.5}

\begin{abstract}
\noindent Traditional network analysis focuses on binary edges, while real-world relationships are more nuanced, encompassing cooperation, neutrality, and conflict. The rise of negative edges in social media discussions spurred interest in analyzing signed interactions, especially in polarized debates. However, the vast data generated by digital networks presents challenges for traditional methods like Stochastic Block Models (SBM) and Exponential Family Random Graph Models (ERGM), particularly due to the homogeneity assumption and global dependence, which become increasingly unrealistic as network size grows. To address this, we propose a novel method that combines the strengths of SBM and ERGM while mitigating their weaknesses by incorporating local dependence based on nonoverlapping blocks. Our approach involves a two-step process: First, decomposing the network into sub-networks using SBM approximation, and, second, estimating parameters using ERGM methods.
We validate our method on large synthetic networks and apply it to a signed Wikipedia network of thousands of editors. Through the use of local dependence, we find patterns consistent with structural balance theory.
\end{abstract}
\spacingset{1.9}


\section{Introduction}
\label{sec:intro}
Network analysis methods are traditionally constrained to binary edges, where relationships are either positive (e.g., alliances, friendships) or negative (e.g., rivalries, conflicts). 
However, these two types of relationships can occur simultaneously in many real-world scenarios. 
Signed networks have been used in diverse contexts, including interstate conflict and cooperation \citep{fritz2022exponential}, social media interactions \citep{leskovec2010signed}, and school networks capturing liking, disliking, and bullying relationships \citep{huitsing2012univariate}.
Historically speaking, signed networks were the basis of one of the first network theories, being theories where the dependence between nodes is embedded into the theory \citep{wassermanSocialNetworkAnalysis1994}:
\citet{heider1946attitudes} proposed the structural balance theory giving rise to the aphorism that ``\textsl{the enemy of my enemy is my friend}''. 
The underlying theory divides triads into balanced and unbalanced. Balanced triads are considered to be more stable and will, therefore, last longer than unbalanced triads.
Recent work has revisited structural balance theory using null models that randomize tie signs while keeping edge positions fixed \citep{gallo2024testing,gallo2025patterns}. Complementary approaches reframe balance as a statistical theory of frustration, treating observed signs as noisy realizations of latent balanced states \citep{gallo2024assessing}
Alternative modeling approaches include decomposing signed networks into separate processes for tie formation and valence assignment \citep{lerner2016structurala}, and entropy-based models that account for constrained network structures \citep{becatti2019entropybased}.

However, the traditional analysis of signed networks was hindered by limited computational resources to estimate more complicated models and by expensive measurements that make their observation possible. 
This situation has changed dramatically with the rise of social media platforms and the availability of digital trace data \citep{lazer2009computational}. 
Online interactions, especially in polarized settings, often contain positive and negative edges. Incorporating negative relations provides a richer understanding of coalitions, conflicts, and divisions within social systems \citep{leskovec2010signed}.

Nevertheless, the modeling of signed networks remains challenging. On the one hand, the class of models specifically designed for signed networks is relatively limited, as most models were developed for unsigned networks. On the other hand, the massive size of modern networks poses separate challenges, as even for unsigned networks many models struggle to scale computationally. At the intersection of these two issues, the need to model both signed relations and large networks, there are little to no methods available.

Two of the most prominent models for network analysis are Stochastic Block Models (SBM, \citealp{nowicki2001estimation}) and Exponential Family Random Graph Models (ERGM, \citealp{lusher2013exponential}).  
An SBM is a latent variable model in which all dependence between edges is accounted for by latent group membership. 
ERGMs, on the other hand, allow for more complex local dependencies through so-called sufficient statistics \citep{denicola2023}. 
Signed network extensions exist for both SBM and ERGM \citep{jiang2015stochastic,yang2017stochastic, fritz2022exponential} with inherent shortcomings:  
While the SBM can represent cohesive subgroups among the nodes, it implies conditional independence for forming edges between nodes. 
This assumption is unrealistic for networks exhibiting more dependence among edges. 
ERGMs, on the other hand, offer a flexible framework to analyze general network properties. 
For a parsimonious model, these properties are commonly assumed to be homogeneous across the entire network. 
The induced global dependence structure, where each edge may technically depend on every other edge in the same way, gives rise to undesirable behavior specifically for large networks \citep{Sc09b,Ha03p}.
Moreover, the estimation process, when based on Markov Chain Monte Carlo (MCMC) methods, often fails due to the complexity and size of the data. 
Larger networks may imply degenerate models with poor sampling properties and longer mixing times \citep{bhamidi2011}.  

In line with \citet{schweinberger2015local}, we fill the resulting methodological gap by combining the strengths of both models while simultaneously addressing their respective weaknesses. 
\citet{schweinberger2015local} introduce the notion of local dependence for binary network, yet the modeling of signed networks in such setting remains an open problem. 
We address this challenge by developing a scalable modeling framework for signed networks under local dependence in Section \ref{sec:model}.  
Our model assumes that each node is only aware of the activities within its block. 
Consequently, the formation of within-block edges can be characterized by a more complex ERGM, while between-block edges are not affected by endogenous effects (i.e., internal and structural influences within a network) that rely on knowledge of neighboring areas. For example, in a social network, individuals may form friendships within their immediate circle without being influenced by the entire network. \citet{snijders2007discussion} considered these models that combine latent space and exponential random graph models as `the next generation of social network models', capable of capturing both cohesive structures and subgraph patterns in larger networks. 

To enable large-scale estimation of our novel model extension, we build on a two-step estimation approach proposed by \citet{babkin2020large}, which we extend to signed networks in Section \ref{sec:est}. 
In the first step, the network is decomposed into sub-networks by approximating the likelihood of the model with an SBM for signed networks. This step is carried out with the help of a variational approximation and computationally fast MM updates \citep{vu2013model}. 
As outlined in Section \ref{sec:uq}, we incorporate uncertainty quantification into this step to assess the reliability of the estimated block structure. 
In the second step, the parameters are estimated given the previously estimated block structure with known methods for ERGMs. 
Thereby, we enable the estimation of signed ERGMs under local dependencies for large signed networks encompassing thousands of nodes. 
The local dependence assumption also speeds up simulation, as the between-block edges are characterized by a less complex model, and enables parallelization, as blocks are conditionally independent given the block membership.

As such, our contributions are: 
\begin{enumerate}
    \item Extension of local dependence from binary to signed networks incorporating size-dependent parametrization in Section \ref{sec:model}. 
    \item Adaption of two-step estimation approach proposed of \citet{babkin2020large} to signed networks in Section \ref{sec:est}.
    \item Introduction of unified uncertainty quantification in Section \ref{sec:uq}, that integrates the uncertainty of decomposing the network into blocks and of estimating the model conditional on the block allocation. 
    \item Application of our model to a large signed network of independent Wikipedia editors in Section \ref{sec:irl}, assessing model fit using out-of-sample cross-validation, which is novel for signed networks.
    \item Development of a open-source R package $\mathtt{bigsergm}$, implementing our full estimation procedure and publicly available at: 
\begin{center}
\url{https://github.com/mschalberger/bigsergm}.
\end{center}
\end{enumerate}

Throughout this paper, we consider a signed network with $N \in \{2, \ldots \}$ nodes. 
The adjacency matrix for this network is denoted as $ \by = (y_{i,j}) \in \mathcal{S}^{N \times N} \coloneqq \mathcal{Y}$, with $ \mathcal{S} \coloneqq \{\!``-",``0",``+"\!\}$. 
The set of all observable signed adjacency matrices among $N$ fixed nodes is defined by $\mathcal{Y}$.
Therefore, $ y_{i,j} = ``-" $ indicates a negative edge, $ y_{i,j} =``+" $ represents a positive edge, and $ y_{i,j} =``0" $ signifies the absence of an edge between nodes $ i $ and $ j $. 
In the undirected case, we have $ y_{i,j} = y_{j,i} $, implying that the adjacency matrix is symmetric. 
For this paper, we assume that the network is undirected and has no self-loops, so $ y_{i,i} =``0" $. 
However, extensions to the directed case follow naturally. 
By $\by_{(-ij)}$, we denote $\by$ excluding $y_{i,j}$.
The set of nodes is partitioned into $ K \in \{2, \ldots \}$ disjoint blocks. Let $N_{k} \in \{2, \ldots \}$ denote the number of nodes in block $k$, for $k=1,\ldots,K$, such that $\sum_{k=1}^{K}N_{k} = N$.
The unobservable vector $ \bz_i = (z_{i1}, \dots, z_{iK}) $ denotes the block membership of each node, where $ z_{i,k} = 1 $ if node $ i $ belongs to block $ k $, and $ z_{i,k} =0 $ otherwise.
Let $\by_{k,l} \in \mathcal{S}^{N_{k} \times N_{l}}$ denote the submatrix of $\by$ that contains all edges between nodes in block $k$ and nodes in block $l$, for $k,l=1,\ldots, K$.
Throughout this paper, capital letters refer to random variables, lowercase letters to their realizations, and bold letters denote vectors or matrices.

\section{Existing Models for Signed Networks} \label{sec:exis}

Statistical models provide a principled framework for analyzing signed networks, with the most prominent examples being ERGMs and SBMs.

\paragraph*{Signed Exponential Random Graph Model.}
The Signed Exponential Random Graph Model (SERGM, \citealp{fritz2022exponential}) posits that an observed network can be fully characterized through a set of sufficient statistics, which capture both endogenous dependencies, which are structural features of the network, and exogenous influences, relating to information such as covariate effects independent of the network \citep{lusher2013exponential}. \citet{fritz2022exponential} defined the probability of observing $\by \in \mathcal{Y}$ by
\be
\label{eq:sergm}
\mbP_{\btheta}(\bY = \by)&=&\dfrac{\exp \left(\btheta^{\top} \bs(\by)\right)}{\kappa(\btheta)},\\
\ee
where $\bs(\by)$ denotes a vector of sufficient statistics, given by a function $\bs: \mathcal{Y} \to \mathbb{R}^p$, weighted by the estimated coefficients $\btheta \in \mathbb{R}^p$ and the normalizing constant $\kappa(\btheta)\, \coloneqq\, \sum_{\tilde{\by} \in \mathcal{Y}} \exp \left(\btheta^{\top} \bs(\tilde{\by})\right)$ guarantees that \eqref{eq:sergm} sums up to 1. 

Interpretation of SERGM parameters is similar to the binary ERGM case. We define positive and negative change statistics, representing the difference in sufficient statistics when changing a specific edge value from the baseline value $``0"$ to a $``+"$ or $``-"$, respectively. Using these, the relative log-odds of observing $Y_{i,j} = ``+"$ or $``-"$ instead of $``0"$ 
\beno \label{eq:logodds}
\log\left( 
\dfrac{\mbP_{\btheta}(Y_{i,j} = ``+" \mid \bY_{(-ij)} = \by_{(-ij)})}
{\mbP_{\btheta}(Y_{i,j} = ``0" \mid \bY_{(-ij)} = \by_{(-ij)})} 
\right)
&=& \btheta^\top \Delta_{i,j}^{0 \rightarrow +}, \s\\
\log\left( \dfrac{ \mbP_{\btheta}(Y_{i,j} = ``-" \mid \bY_{(-ij)} = \by_{(-ij)})}
{\mbP_{\btheta}(Y_{i,j} = ``0" \mid \bY_{(-ij)} = \by_{(-ij)})} \right)
&=& \btheta^\top \Delta_{i,j}^{0 \rightarrow -}.
\ee
Each component of $\btheta$ represents the change in the log-odds of observing a positive or negative edge instead of no edge that is associated with a one-unit increase in the corresponding change statistic. Therefore, a positive estimate increases the relative log-odds of $Y_{i,j} = ``+"$ or $Y_{i,j} =``-"$ compared to $Y_{i,j} =``0"$, while a negative estimate decreases these log-odds.

When facing large networks, this framework faces several challenges: 
First, evaluating the normalization constant $\kappa(\btheta)$ is intractable for almost all networks.
Statistical inference generally relies on a MCMC approximations of the likelihood function, which often fails due to the complexity and size of the large networks.
Second, the sufficient statistics induce global dependence by being calculated over the entire network. 
The inclusion of these statistics, therefore, implies knowledge about the entire network, which is unrealistic for networks with more than a few thousand nodes.

\paragraph*{Signed Stochastic Block Model.}
One approach to restrict the nodes' knowledge locally is to cluster into blocks. 
A seemingly simple model along these lines assumes that nodes exhibit structural equivalence within their respective block \citep{fienberg1981}. 
Assuming that the block assignments are latent random variables that we learn from observed networks leads to the Stochastic Block Model (SBM), where the probability of any observed network $\by \in \mathcal{Y}$ given a latent clustering $\bZ$ is:
\beno
\label{eq:sbm}
\mbP_{\bm{\pi}}(\bY = \by\mid \bZ = \bz)&=& \dsum_{i<j} \mbP_{\bm{\pi}}(Y_{i,j} = y_{i,j}\mid Z_i = k, Z_j = l),
\ee
where the sum ranges over all unordered pairs of distinct nodes $(i, j)$ in the network.
The conditional probability to observe $Y_{i,j}=y_{i,j}$ is assumed to be the probability mass function of a Bernoulli-distributed random variable:
\be
\label{eq:conditional_probability}
\mbP_{\bm{\pi}}(Y_{i,j} = y_{i,j}\mid Z_i = k, Z_j = l) &=& \pi_{k,l}^{y_{i,j}}\, (1-\pi_{k,l})^{1-y_{i,j}},
\ee
where $\pi_{k,l} \in [0,1]$ is the block-specific connection probability between blocks  $k$  and  $l$.

By substituting the conditional Bernoulli distribution in \eqref{eq:conditional_probability} with a multinomial distribution,  we can accommodate for signed edges $y_{i,j}\in \mathcal{S}$: 
\beno
\mbP_{\bm{\pi}}(Y_{i,j} = y_{i,j}\mid Z_i = k, Z_j = l) &=& \dprod_{y \in \mathcal{S}}\pi_{k,l}(y)^{\mI(y_{i,j} = y)} 
\ee
where $\pi_{k,l}(y)$ with $y\in \mathcal{S}$ denotes the probability of $Y_{i,j} = y$ with
$\sum_{y \, \in\,\mathcal{S}}\pi_{k,l}(y) = 1$ and ${\bm{\pi}} = (\pi_{k,l}(y))$ \citep{li2023ssbm}.
A shortcoming of this model is its simplicity: The probability to observe a negative, neutral, or positive edge between nodes $i$ and $j$ solely depends on the block memberships of the involved nodes. This dependence structure is not able to capture foundational theories about signed networks, such as the structural balance theory \citep{heider1946attitudes}.

\section{Exponential Random Graph Model for Signed Networks under Local Dependence} \label{sec:model}

In many large networks, edge dependencies are local rather than global: nodes are unlikely to be aware of all others' ties, making global dependence assumptions implausible. We therefore restrict dependence to disjoint subpopulations of nodes, or blocks. This approach follows earlier work on binary networks that models local dependence via latent block structures \citep{schweinberger2015local}. 
However, extending this idea to non-binary data, such as signed networks, is non-trivial. 

We model local dependence through a latent block structure $\bZ$, which groups nodes into unobserved communities that govern local interaction patterns. Given this block structure, the probability of observing a network $\by$ decomposes into two components: one capturing complex dependencies within blocks, and another describing between-block interactions assumed to be independent. This decomposition reflects modular structure commonly observed in real networks and enables efficient, parallel computation. This yields the following factorization for network $\by \in \mathcal{Y}$:

\be 
\label{eq:decomp}
    \mbP_{\btheta}\left(\bY = \by \mid \bZ = \bz\right) &=&
    \left(\dprod_{k=1}^{K}\mbP_{\btheta_{k,k}}\left(\bY_{k,k}=\by_{k,k}\mid \bZ = \bz\right)\right) \s\\
    &\times& \left(
    \dprod_{k<l}\mbP_{\btheta_{k,l}}\left(\bY_{k,l}=\by_{k,l}\mid \bZ =\bz\right) \right),
\ee
where $\btheta \coloneqq \operatorname{vec}(\btheta_{1,1}, \ldots , \btheta_{K,K},\btheta_{1,2}, \ldots, \btheta_{K-1,K})$ denotes the vector of both between-block and within-block parameters with $\operatorname{vec}$ defining a function that stacks its arguments vertically. 
Unlike with stochastic block models, this model does not assume that edges within and between blocks are independent. Rather, only the independence of between-block edges is guaranteed, while within-block edges can be strongly dependent.

For nodes within the same block $k$, the probability of observing the sub-network $\by_{k,k}$ is given by a complex signed ERGM:
\be
\label{eq:within}
    \mbP_{\btheta_{k,k}}\left(\bY_{k,k} = \by_{k,k}\mid \bZ = \bz\right) &=& \dfrac{\exp(\btheta_{k,k}^\top \, \bs(\by_{k,k}))}{\kappa(\btheta_{k,k})},
\ee
where $\btheta_{k,k} \in \mathbb{R}^{p}$  denotes the vector of within-block parameters for block $k$, and $\bs: \mathcal{Y}_{k,k} \rightarrow \mathbb{R}^{p}$ is a function calculating the sufficient statistics. 
This function includes both dyad-independent and dyad-dependent variables, since we assume that nodes are aware of the activities within their block. 
The number of sufficient statistics used to model within-block networks $p \in \{1, \ldots \}$ is assumed to be constant across blocks.

The between-block edges are characterized by a signed SBM, where the probability of observing the sub-network $\by_{k,l}$, conditional on the block assignments $\bz$, is given by:
\beno
\label{eq:between}
    \mbP_{\btheta_{k,l}}\left(\bY_{k,l} = \by_{k,l}\mid \bZ = \bz\right) &=&\dfrac{\exp\left(\btheta_{k,l}^\top \bh(\by_{k,l})\right)}{\kappa(\btheta_{k,l})},
\ee
with $\btheta_{k,l} \in \mathbb{R}^{q}$ denoting the between-block parameters.
The sufficient statistics $h: \mathcal{Y}_{k,l} \to \mathbb{R}^{q}$ (with $l \neq k$) include only dyad-independent variables, since dyad-dependency would require knowledge of the activity in other blocks: 
\beno
\bh_{k,l}(\by_{k,l})&=&\dsum_{i,j}z_{i,k}z_{j,l}\left(\mI(y_{i,j}=``-")\bx_{i,j,-} + \mI(y_{i,j}=``+")\bx_{i,j,+} \right),
\ee
where $\bx_{i,j,+}$ and $\bx_{i,j,-}$ are covariate vectors associated with positive and negative edges, respectively.
Since this sub-model implies dyadic independence, the normalizing constant has a closed form:
\beno
\kappa_{k,l}(\btheta_{k,l}) &=& \dprod_{i,j:z_{i,k}z_{j,l}=1} \left(1 + \exp(\bx_{i,j,+}^{\top}\btheta_{k,l,+}) 
+ \exp(\bx_{i,j,-}^{\top}\btheta_{k,l,-}) \right),
\ee
where $\btheta_{k,l,+}$ and $\btheta_{k,l,-}$ are the corresponding coefficient vectors for block pair $(k,l)$.
Technically, one could also include dependence-inducing statistics in this sub-model. However, this would relax the local dependence assumption across blocks and undermine the asymptotic arguments underlying the model, including the justification of the two-step estimation approach.
In practice, we are interested in parsimonious models to ease interpretation, hence letting $\btheta_{k,l}$ vary freely over $k,l = 1, \ldots, K$ is unreasonable. 
Extending \citet{krivitsky2023tale}, we represent different homogeneity assumptions on these coefficients by linear combinations of population-level coefficients for within and between blocks, $\bbeta_w  \in \mathbb{R}^{s\times p}$ and $\bbeta_b\in \mathbb{R}^{t\times q}$, and block-specific covariates, $\bv_k \in \mathbb{R}^s$ and $\bu_{k,l} \in \mathbb{R}^t$:
\be
\label{eq:lp}
\btheta_{k,k} &\coloneqq& (\bv_{k}^\top\bbeta_w)^{\top} \text{ and }\btheta_{k,l} &\coloneqq& (\bu_{k,l}^\top\bbeta_b)^{\top}.
\ee
The model would constitute a curved ERGM with local dependence if the relationship between the population-level coefficients and the block-specific covariates in equation \eqref{eq:lp} were non-linear \citep{hunter2006inference}.
Block-specific covariates $\bv_{k}$, such as block size $N_{k}$ or $\log(N_{k})$, further enable size-dependent parametrizations (see, e.g., \citealp{butts2015}).
Larger blocks will typically have a lower density, however using non-size dependent parameters would lead to preservation of the density.
Let $\beta_{w,a,b}$, $\theta_{k,k,a}$ , and $v_{k,b}$ denote the $(a,b)$th entry of $\bbeta_w$, the within-block effect of the $a$th statistic in $\btheta_{k,k}$, and value of the $b$th entry of $\bv_i$, respectively. 
Then $\theta_{k,k,b}$ is given by $\beta_{w,a,b}\,\bv_{k}$, and conditional on $\bz$, $\beta_{w,a,b}$ can be interpreted analogously to the basic SERGM in Section \ref{sec:exis} as the increase of $\theta_{k,k,b}$ associated with a one-unit increase in $v_{k,b}$.
Covariates $\bu_{k,l}$ typically only include a constant, dummy variables for specific pairs of blocks, or functions of $\bv_{k}$ and $\bv_{l}$. 
\begin{example}[Signed SBM]
\label{par:ex1} 
To start edges $y_{i,j}$ are assumed to be conditionally independent given the block memberships in Section \ref{sec:exis}.    
Let $\mI(y_{i,j} = y) = a_{i,j,y}$ be an indicator for observing edge value $y \in \{ ``+", ``-"\}$ between node $i$ and $j$.
The vector of sufficient statistics for the within-block model $\bs(\by_{k,k}) \in \mathbb{R}^{2}$ and the between-block model  $\bh(\by_{k,l}) \in \mathbb{R}^{2}$ then include the count of positive and negative edges for each block or block pair, respectively. Define the edge terms as: 
\beno 
\text{Edges}^+(\by) &\coloneqq& \dsum_{i < j} a_{i,j,+}, \quad
\text{Edges}^-(\by) &\coloneqq& \dsum_{i < j} a_{i,j,-}.
\ee
Then the vectors of sufficient statistics is given by:
\beno
\bs(\by_{k,k})  &=& \bh(\by_{k,l}) &=& 
\left( \text{Edges}^+(\cdot), \,
\text{Edges}^-(\cdot)\right)^\top.
\ee
These statistics correspond to parameters $\btheta_{k,l} = (\theta_{k,l,+}, \theta_{k,l,-})$, where we define $\theta_{k,l,0}= 0$. 
The multinomial probabilities from \eqref{eq:conditional_probability} are then:
\beno 
\pi_{k,l}(y) &=& \dfrac{\exp(\theta_{k,l,y})}
{\dsum_{y^\star \in \mathcal{S}} \exp(\theta_{k,l,y^\star})}.
\ee
Thus, $\theta_{k,l,+}$ and $\theta_{k,l,-}$ represent log-odds of observing a positive and negative edge between blocks $k$ and $l$ compared to observing no edge. 
In this form, each $\bs(\by_{k,k})$ and $\bh(\by_{k,l})$ acts like an intercept term: it aggregates the number of edges of a particular sign, and its corresponding parameter controls the baseline probability of that edge type appearing between nodes with the given block memberships.
For the within-block model, we can also represent this in the linear predictor structure from Equation \eqref{eq:lp}. In this case, each $\bv_{k} \in \mbR^K$ is a one-hot vector with a 1 in the $k$th position and 0 elsewhere.
Similarly, for the between-block model, $\bu_{k,l} \in \mbR^{K}$ is a vector with 1s in position $k$ and $l$.
To reduce model complexity and avoid estimating block-pair-specific parameters, we assume homogeneous parameters across all between-block pairs. In this case, we define $\bu_{k,l} = 1$, so all $\btheta_{k,l} = \bbeta_b \in \mbR^{1\times 2}$
In the following examples, this parameter vector remains fixed, as no dyad-independent terms are added to the between-block model.
\end{example}

\begin{example}[Signed Model with Triadic Terms]
\label{par:ex2} 
To relax the assumption of the SSBM that edges within and between blocks are independent, we can incorporate triadic terms that reflect structural balance theory.  Specifically, the model may include triadic configurations reflecting the principles ``\textsl{friend of my friend is my friend}'' and ``\textsl{enemy of my enemy is my friend}''   \citep{fritz2022exponential}.
However, the inclusion of triadic terms often leads to model degeneracy \citep{Sc09b}. 
\citet{ScSt19} showed that defining triadic statistics by: 
\beno
    \text{CF}^{+}(\by_{k,k}) &\coloneqq& \dsum_{i < j}  a_{i,j,+} \, \mI \left(\dsum_{h \neq i,j} a_{i,h,+} \,  a_{h,j,+} > 0\right) , \s \\ 
    \text{CE}^{+}(\by_{k,k}) &\coloneqq& \dsum_{i < j} a_{i,j,+} \, \mI\left(\dsum_{h \neq i,j} a_{i,h,-} \, a_{h,j,-} > 0\right)
\ee
yields better-behaved distributions.
The vector of sufficient statistics for the within-block model is given by:
\beno
    \bs(\by_{k,k}) &=& \left(
        \text{Edges}^+(\by_{k,k}), \,
        \text{Edges}^-(\by_{k,k}) , \,
        \text{CF}^{+}(\by_{k,k}) , \, 
         \text{CE}^{+}(\by_{k,k})
    \right)^\top,
\ee
where the triadic terms count the number of positive edges that are connected through at least one common friend, in the case of $\text{CF}^{+}(\by_{k,k})$, or at least one common enemy, in the case of $\text{CE}^{+}(\by_{k,k})$.
\end{example}

\begin{example}[Signed Model with Structural Terms]
\label{ex:3}

To capture complex local structures beyond simple count statistics such as edge or triad counts, we incorporate geometrically weighted degree and edgewise shared partner statistics, including a geometrically weighted version of the \textsl{``enemy of my enemy is my friend"} triad configuration. These statistics account for variation in local connectivity patterns, particularly in networks characterized by degree heterogeneity and a tendency toward triadic closure.
By down-weighting high-degree nodes and configurations with many shared partners or degrees, geometric weighting reduces the influence of extreme values and improves the stability of parameter estimation. 
The decay parameter $\omega$ controls this down-weighting: smaller values of $\omega$ place more weight on high degree nodes or triads with many shared partners, while larger values of $\omega$ reduce their influence.

The geometrically weighted degree statistic for positive or negative edges within block $k$ is defined as:
\beno
\text{GWD}^{y}(\by_{k,k}, \omega)  &\coloneqq& \exp(\omega) \dsum_{d=1}^{N_k - 1} \left( 1 - (1 - \exp(-\omega))^d \right) \text{deg}_{k,d}^y,
\ee
where
\beno
\text{deg}_{k,d}^y &\coloneqq& \dsum_{i=1}^{N_k} \mI\left(\dsum_{j \neq i} a_{i,j,y} = d \right)
\ee
counts nodes in block $k$ with a positive or negative degree of $d$.
Analogously, the geometrically weighted edgewise shared partner statistic is given by:
\beno
\text{GWESE}^{y}(\by_{k,k}, \omega) &\coloneqq& \exp(\omega) \dsum_{d=1}^{N_k - 2} \left( 1 - (1 - \exp(-\omega))^d \right)  \text{ESE}_{d}^y, \s\\ 
\text{GWESF}^{y}(\by_{k,k}, \omega) &\coloneqq& \exp(\omega) \dsum_{d=1}^{N_k - 2} \left( 1 - (1 - \exp(-\omega))^d \right)  \text{ESF}_{d}^y, \\
\ee
where
\beno
\text{ESE}_{d}^y \,\coloneqq\, \dsum_{i<j} a_{i,j,y} \, \mI\left( \dsum_{h \neq i,j} a_{i,h,-} \,a_{j,h,-} = d \right), \quad
\text{ESF}_{d}^y \,\coloneqq\, \dsum_{i<j} a_{i,j,y}  \,\mI\left( \dsum_{h \neq i,j} a_{i,h,+} a_{j,h,+} = d \right).
\ee
Here, $\text{ESE}_{d}^y$ counts edges of sign $y$ sharing exactly $d$ common enemies, and $\text{ESF}_{d}^y$ counts edges of sign $y$ sharing exactly $d$ common friends.
The vector of sufficient statistics for the within-block model is then defined as: 
\beno
\bs(\by_{k,k}) &=& (
 \text{Edges}^+(\by_{k,k}), \,
        \text{Edges}^-(\by_{k,k}), \,
    \text{GWD}^+(\by_{k,k}, \omega) , \\
    &~&\text{GWD}^-(\by_{k,k}, \omega) , \,
    \text{GWESE}^+(\by_{k,k}, \omega) 
    )^\top.
\ee
\end{example}
\section{Scalable Estimation}
\label{sec:est}

The block allocation  $\bZ$, a multinomial variable taking values in $\{1, \ldots, K\}$, is often unobserved and must be learned from the observed network $\by$. 
Assuming that the number of blocks $K$ is fixed, we model $\bZ$ under a multinomial prior with relative group-sizes $\gamma_1, \dots, \gamma_K$:
\be
\label{eq:prior}
 Z_i &\overset{\text{iid}}{\sim}& \text{Multinomial}(1;\gamma_1, \dots, \gamma_K),
\ee
following the practice for SSBMs \citep{li2023ssbm}.
 Combining \eqref{eq:prior} with the conditional model $\bY \mid \bZ = \bz$ defined in \eqref{eq:decomp} yields a latent variable model (see, \citealp{david2011}, for an introduction to the field). 
 The log-likelihood for parameters $\btheta$ and $\bgamma$ is
 \be
\label{eq:likelihood}
\ell(\btheta, \bgamma) &=& \log \left(\mbP_{\btheta, \bgamma}(\bY = \by)\right) &=& 
    \log \left(\dsum_{\bz \in \mathcal{Z}}\mbP_{\btheta}(\bY = \by \mid \bZ = \bz)\, 
    \mbP_{\bgamma}(\bZ =\bz)\right)
\ee
where $\mathcal{Z} \coloneqq \{1, \ldots, K\}^N$ is the discrete space of block assignments.

Plugging \eqref{eq:decomp} into $\ell(\btheta, \bgamma)$, shows that direct maximization of \eqref{eq:likelihood} would involve the evaluation of nested intractable sums: the normalization constants of the within networks \eqref{eq:within} and the discrete space of block assignments, both growing superexponentially with $N$.

Thus, we introduce an auxiliary distribution $q$ over $\mathcal{Z}$
and apply Jensen's inequality:
\be
\label{eq:elbo}
\ell(\btheta, \bgamma) &\geq& 
\mathbb{E}_q\left(\log\left(\mbP_{\btheta}(\bY = \by \mid \bZ = \bz)\, \mbP_{\bgamma}(\bZ = \bz)\right)\right) - \mathbb{E}_q(\log q(\bZ)) \\&\eqqcolon& \mathrm{LB}(q, \btheta, \bgamma),
\ee 
where $\mathrm{LB}(q, \btheta, \bgamma)$ is a functional of $q, \btheta, \bgamma$ called the evidence lower bound.
 The expectation under $q$ over a function $f$ is $\mathbb{E}_q(f(\bZ)) \coloneqq \sum_{z \in \mathcal{Z}} q(\bz) f(\bz)$.  
We optimize $\mathrm{LB}(q, \btheta, \bgamma)$ regarding $q,\btheta,$ and $\bgamma$ by alternating block coordinate ascent:
\begin{flalign} \label{alg:optimization_steps}
\begin{aligned}
    &\textbf{Step 1: } \text{Update } q \text{ to maximize } \mathrm{LB}(q, \btheta, \bgamma) \text{ for fixed } \btheta \text{ and } \bgamma. \hspace{3.1cm} \\
    &\textbf{Step 2: } \text{Update } \btheta \text{ and } \bgamma \text{ to maximize } \mathrm{LB}(q, \btheta, \bgamma) \text{ for fixed } q.
\end{aligned}
\end{flalign}
If $\bz$ is observed or can be derived from covariate information, only Step 2 is needed. 

The optimal solution in Step 1 is $q(\bZ) =  \mathbb{P}(\bZ = \bz \mid \bY = \by)$, which does not admit a closed-form expression \citep{matias2014}. 
For SBMs of moderate size, Gibbs sampling based on the full conditional distributions $Z_i \mid Z_1, ..., Z_{i-1}, Z_{i+1}, ..., Z_{N}$ can be employed
\citep{nowicki2001estimation}, 
but for larger networks, as considered in this paper, variational methods offer scalable approximations \citep{BlKuMc17}.
For this approach, we restrict $q$ to a tractable family of distributions parametrized by $\balpha$ and approximate $\mathbb{P}(\bZ = \bz \mid \bY = \by) \approx q_{\balpha}(\bz)$, where $q_{\balpha}(\bz)$ minimizes the Kullback-Leibler divergence to $\mathbb{P}(\bZ = \bz \mid \bY = \by)$.
An explicit form of this variational family is provided in Section \ref{sec:variational}.
Still, the dependence of within-block networks \eqref{eq:within} inhibits adaption to our setting. 
In Section \ref{sec:approx}, we adapt the result of \citet{babkin2020large} to signed networks showing that under suitable conditions the probability \eqref{eq:decomp} can be approximated by the probability of an SSBM, mentioned in Section \ref{sec:exis} and Example \ref{par:ex1}. 
This result justifies the use of scalable optimization methods originally developed for SBMs and decouples the updates of $q$ and $\bgamma$ from the updates of $\btheta$.
Thus, the alternating block coordinate ascent algorithm detailed in \eqref{alg:optimization_steps} becomes a two-step algorithm without the need to iterate between the two steps:  
\begin{flalign} 
\label{eq:algorithm2}
\begin{aligned}
    &\textbf{Step 1: } \text{Update } q \text{ and } \bgamma \text{ to maximize } \mathrm{LB}(q, \btheta, \bgamma) \text{ via SSBM approximation.}\hspace{0.9cm} \\
    &\textbf{Step 2: } \text{Update } \btheta \text{ to maximize } \mathrm{LB}(q, \btheta, \bgamma) \text{ for fixed } q \text{ and } \bgamma.
\end{aligned}
\end{flalign}
In Section \ref{sec:variational}, we describe Step 1 and adapt the approach of \citet{vu2013model} based on MM Steps to the signed setting. 
Section \ref{sec:conditional_the} introduces a scalable algorithm for Step 2.
\hide{
, we first approximate the intractable likelihood resulting from \eqref{eq:decomp} by the pseudo likelihood, which was already studied by \citet{stewart2025pseudolikelihoodbased}. 
Then we propose two related methods to maximize \eqref{eq:elbo} inspired by \citet{WeTa90}: 
\begin{enumerate}
    \item Set $\hat{\bz} = (\argmax\limits_{k \, \in\, \{1, \ldots, K\}} \alpha_{i,k})_{i=1}^N$ and estimate $\btheta$ conditional on this block assignment, like an EM-type algorithm.
    \item Sample $T$ times from the approximate posterior, estimate conditional on all samples block assignments, and average the results, like an MCMC approximation of \eqref{eq:elbo}.
\end{enumerate}
Section \ref{sec:conditional_the} discusses the estimation of $\btheta$ conditional on $\bZ = \hat{\bz}$, which is necessary in both options.
}
The uncertainty of this algorithm is quantified in Section \ref{sec:uq}.

\subsection{Signed Stochastic Block Model Approximation}
\label{sec:approx}

For any model specification, we recover the Signed Stochastic Block Model (SSBM), as shown in Example \ref{par:ex1}, by setting all parameters in model \eqref{eq:within} corresponding to dyad-dependent statistics (e.g., transitivity) to zero.
Note that the sufficient statistics in \eqref{eq:between} are already dyad-independent by definition. 
Therefore, the approximation of the probability of our model by the probability of an SSBM only accrues errors for within-block networks. 
As $N$ and $K$ grow, our joint probability will then increasingly be dominated by terms relating to between-block probabilities, 
provided single blocks are not too large.
This offers a heuristic justification for why the SSBM provides a good approximation in large populations.

Accordingly, we decompose the within-block parameters of the $k$th block \newline  $\btheta_{k,k} \,=\, \operatorname{vec}(\btheta_{k,k}^{\not\perp}, \btheta_{k,k}^{ \perp})$, where $\btheta_{k,k}^{\not\perp}$ relates to statistics inducing dependence and $\btheta_{k,k}^{\perp}$ to statistics implying independence.  
Setting $\btheta_{k,k}^{\not\perp} = \bm{0}$, we define $\btheta_{k,k}^{\text{SBM}} = \operatorname{vec}(\bm{0}, \btheta_{k,k}^{\perp})$ for all $k \in \{1, \ldots, K\}$ and $\btheta^{SBM} \coloneqq \operatorname{vec}(\btheta_{1,1}^{\text{SBM}}, \ldots , \btheta_{K,K}^{\text{SBM}},\btheta_{1,2}, \ldots, \btheta_{K-1,K})$.
With this notation, substituting $\btheta$ by $\btheta^{SBM}$ in \eqref{eq:decomp}, defines the corresponding nested SSBM. 
Denote by $m(\bz)$ the size of the largest block in block allocation  $(\bz) \in \mathcal{Z}$ and let $d: \mathcal{Y} \rightarrow \mathcal{Y}$ be the Hamming distance between two signed networks $\by_{1}, \by_{2} \in \mathcal{Y}$:
\beno
d\left(\by_{1}, \by_{2}\right)&=&\dsum_{i<j} \mI(\by_{1, i, j} \neq \by_{2, i, j}),
\ee
 where $\mI$ denotes the indicator function.
We require the following conditions:
\begin{assumption}[: Smoothness of $\bs$]
\label{eq:assumption1}
 A constant $c > 0$ exists such that for all $\btheta \in \bTheta$ and all $\by_1, \by_2 \in \mathcal{Y}$, 
     \beno 
          \left| \langle \btheta, \bs(\by_1) - \bs(\by_2) \rangle \right| &\leq& c \, d(\by_1, \by_2) \, m(\bz) \log N.
          \ee
\end{assumption}
\begin{assumption}[: Smoothness of $\btheta$]
\label{eq:assumption2}
A constant $c > 0$ exists such that for all $\btheta_{k,l;1}, \btheta_{k,l;2} \in \bTheta_{k,l}$ and all $\bz \in \mathbb{Z}$,
    \beno 
         \lvert\langle\btheta_{k, l; 1}-\btheta_{k, l; 2}, \mathbb{E}(\bs_{k,l}(\bY))\rangle\rvert 
         &\leq& c\, \lvert\btheta_{k, l; 1}-\btheta_{k, l; 2}\rvert m(\bz)^{2} \log N.
         \ee
\end{assumption}

All sufficient statistics considered in this paper satisfy Assumptions \ref{eq:assumption1} and \ref{eq:assumption2}.
Following Theorem 2 from \citet{babkin2020large}, we get:
\be
\label{eq:approx}
   \mbP_{\btheta}(\bY = \by\mid \bZ = \bz) &\approx& \mbP_{\btheta^{\text{SBM}}}(\bY = \by\mid \bZ = \bz).
\ee

\subsection{Variational Approximation based on MM Updates}
\label{sec:variational}

Supported by this result, we approximate $\mbP_{\btheta}(\bY = \by \mid \bZ = \bz)$ in  \eqref{eq:elbo} by the probability of an SSBM, which simplifies the optimization in Steps 1 and 2 outlined above.
We assume a mean-field approximation of $\mathbb{P}(\bZ = \bz \mid \bY = \by)$, implying that it factorizes across nodes:
\beno
    q_{\balpha}(\bz)&=& \dprod_{i=1}^N q_{\balpha_i, i}(\bz_i),
\ee 
where $q_{\balpha_i, i}$ denotes the probability distribution of a categorical distribution parametrized by $\balpha_i = (\alpha_{i,k})\in [0,1]^K$ with $\sum_{k =1}^K \alpha_{i,k} = 1$.
The variational parameters that characterize the class of distribution approximating  $ \mathbb{P}(\bZ = \bz \mid \bY = \by)$
are $\balpha = (\balpha_i) \in [0,1]^{K \times N}$.

The lower bound $\mathrm{LB}(q, \btheta, \bgamma)$ defined in \eqref{eq:elbo} is then a function of $\balpha$:
\beno
\mathrm{LB}(\balpha, \btheta, \bgamma) &=&\dsum_{i<j}\dsum_{k=1}^K\dsum_{l=1}^K \alpha_{i,k}\alpha_{j,l}\log p_{k,l}(y_{i,j} )+\dsum_{i=1}^N\dsum_{k=1}^K\alpha_{i,k}(\log\gamma_k-\log\alpha_{i,k}),
\ee
where $p_{k,l}(y) = \mathbb{P}_{\btheta^{\text{SBM}}}(Y_{i,j} = y \mid Z_i =k, Z_j = l)$ is the multinomial probability distribution evaluated at $y$ of an edge between nodes $i$ and $j$, with $Z_i =k$ and $ Z_j = l$ under the SSBM approximation described in Section \ref{sec:approx}.
Note that $p_{k,l}(y)$ depends on \(\btheta^{\mathrm{SBM}}\). 

Although one can directly maximize the lower bound \(\mathrm{LB}(\balpha, \btheta, \bgamma)\) with respect to \(\balpha\) using iterative fixed-point methods (see, e.g., \citealp{DaPiRo08}), a more scalable and robust approach employs Minorization-Maximization (MM) steps \citep{vu2013model}.
Specifically, we introduce the surrogate function $Q({\bgamma}^{(t)},{\btheta}^{(t)};{\balpha}^{(t)},{\balpha})$:
\be
\label{eq:minorizer}
    Q({\bgamma}^{(t)},{\btheta}^{(t)};{\balpha}^{(t)},{\balpha})  &=& \dsum_{i=1}^N \dsum_{k=1}^K A_{i,k}(\by,\balpha^{(t)}, {\btheta}^{(t)})\alpha_{i,k}^2 + 
    B_{i,k}({\bgamma}^{(t)}, \balpha^{(t)})\alpha_{i,k}
\ee
 which is quadratic in $\balpha$ and hence easier to maximize.  
The quadratic term is defined as 
\beno
    A_{i,k}(\by,\balpha^{(t)}, {\btheta}^{(t)}) &\coloneqq& \dfrac{\Omega_{i,k}^{(t)}(\by,\balpha^{(t)}, {\btheta}^{(t)})}{ \alpha_{i,k}^{(t)}} - \dfrac{1}{\alpha_{i,k}^{(t)}},
\ee
and the linear term  as
\beno
     B_{i,k}({\bgamma}^{(t)}, \balpha^{(t)}) &\coloneqq& \log\gamma_k^{(t)}-\log\alpha_{i,k}^{(t)} +1,
\ee
where 
\be
\label{eq:omega_main}
    \Omega_{i,k}^{(t)}(\by,\balpha^{(t)}, {\btheta}^{(t)}) &\coloneqq& \dsum_{j\neq i}^N\sum_{l=1}^K \alpha_{j,l}^{(t)}\log\, p_{k,l}^{(t)}(y_{i,j}).
\ee
This surrogate function minorizes $\mathrm{LB(\balpha, \btheta, \bgamma)}$ in $\balpha^{(t)}$ and abides the following properties:
\beno
Q(\bgamma^{(t)},\btheta^{(t)},\balpha^{(t)};\balpha)&\le&\mathrm{LB}(\balpha, \bgamma^{(t)},\btheta^{(t)}) \text{~~~for all }\balpha \in [0,1]^{K\times N}\\
Q(\bgamma^{(t)},\btheta^{(t)},\balpha^{(t)};\balpha^{(t)})&=&\mathrm{LB}(\balpha^{(t)},\bgamma^{(t)},\btheta^{(t)}).
\ee
Increasing $Q(\bgamma^{(t)}, \btheta^{(t)}; \balpha^{(t)}, \balpha)$ with respect to $\balpha$ guarantees an increase in $\mathrm{LB}(\balpha, \btheta, \bgamma)$. 
Since $Q$ is a quadratic function of $\balpha$,  scalable and robust optimization methods are available to efficiently perform Step 1.
The direct evaluation of \eqref{eq:omega_main} presents a computational bottleneck with $\mathcal{O}(N^2K^2)$ complexity. 
To mitigate this, we reformulate the calculation of \eqref{eq:omega_main} across all $i, k$ as a matrix product, facilitating the use of accelerated sparse matrix operations (see Supplement \ref{sec:sparse} for the full derivation).
Further information on the derivation of Equation \eqref{eq:minorizer} and the updates of $\btheta^{\text{SBM}}$ and $\bgamma$ are provided in Supplement \ref{sec:comp}.


\subsection{Estimation with known Blocks}
\label{sec:conditional_the}

Running the algorithm from the previous section until convergence provides us with an estimate of the posterior distributions and $\bgamma$. 
Given $\btheta^{\text{SBM}}\neq  \btheta$, we estimate $\btheta$ as detailed in Step 2 from \eqref{eq:algorithm2}. 
We approximate the intractable likelihood resulting from \eqref{eq:decomp} by the pseudo likelihood, which was already studied by \citet{stewart2025pseudolikelihoodbased}. 
We propose two methods to maximize \eqref{eq:elbo},  inspired by \citet{WeTa90}: 
\begin{enumerate}
    \item Set $\hat{\bz} = (\argmax\limits_{k \, \in\, \{1, \ldots, K\}} \alpha_{i,k})_{i=1}^N$ and estimate $\btheta$ conditional on this block assignment, like an EM-type algorithm similar to \citet{babkin2020large}.
    \item Sample $T$ times from the approximate posterior, estimate $\btheta$ conditional on all samples block assignments, and average the results, like an MCMC approximation of \eqref{eq:elbo}.
\end{enumerate}
As shown in Sections \ref{sec:sim} and \ref{sec:irl}, both lead to very similar results. 
In what follows, we discuss the estimation of $\btheta$ conditional on $\bZ = \hat{\bz}$, which is completes the description of both options. 

 Given the block membership vector $\bz$, we can estimate the model parameters $\btheta$. 
 Since \eqref{eq:lp} expresses $\btheta$ as a function of the population-level parameters $\bbeta = (\bbeta_w, \bbeta_b)$, we optimize over these lower-dimensional parameters rather than the full set of block- or block-pair-specific coefficients $\btheta_{k,k}$ and $\btheta_{k,l}$.
 Accordingly, we reparametrize \eqref{eq:decomp} using the coefficients $\bbeta_{w,\operatorname{vec}} \coloneqq \operatorname{vec}(\bbeta_w)$ and $\bbeta_{b,\operatorname{vec}} \coloneqq\operatorname{vec}(\bbeta_b)$, together with the block-specific sufficient statistics 
\be
\label{eq:reparam_suff}
\bs_k \colon \by_{k,k} \mapsto \bv_k \otimes \bs(\by_{k,k}), \qquad
\bh_{k,l} \colon \by_{k,l} \mapsto \bu_{k,l} \otimes \bh(\by_{k,l}).
\ee
The term $\bm{a} \otimes \bm{b}$ denotes the Kronecker product of vectors $\bm{a} \in \mathbb{R}^m$ and $\bm{b} \in \mathbb{R}^n$. 
In the following, we focus on estimating the within-block model parameter $\bbeta_{w,\operatorname{vec}} $. 
 The method applies analogously to $\bbeta_{b,\operatorname{vec}}$.
 While both MCMC-based maximum likelihood (MLE,\citealp{hunter2006inference}) and maximum pseudo-likelihood methods (MPLE, \citealp{stewart2025pseudolikelihoodbased}) can be used to estimate the parameters in the second step, MPLE is generally better suited for large networks due to its computational efficiency and scalability. If dyadic independence holds, both approaches coincide.

The conditional probability of an edge between nodes $i$ and $j$ with $z_i = z_j = k$ to be $y \in \mathcal{S}$ given the rest of the network $\bY_{(-ij)}$ and $\bbeta_{w,\operatorname{vec}}$ is
\beno
    \mbP_{\bbeta_{w,\operatorname{vec}}}(Y_{i,j} = y \mid \bY_{(-ij)} = \by_{(-ij)}, z_i = z_j = k) &=& \dfrac{\exp \left(\bbeta_{w,\operatorname{vec}}^{\top} \bs_k(\by_{i,j}^y)\right)}{\dsum_{y^*} \exp \left( \bbeta_{w,\operatorname{vec}}^{\top} \bs_k(\by_{i,j}^{y^*}) \right)}.
\ee
Here, $\bs_k$ from \eqref{eq:reparam_suff} is evaluated at the network configuration $\by_{i,j}^y$, where the value of edge $y_{i,j}$ is set to $y$, while all other edge values remain fixed as observed.

 With $\Delta_{i,j,k}^{0 \rightarrow +}(\by)$ and $\Delta_{i,j,k}^{0 \rightarrow -}(\by)$ being the change in the reparametrized sufficient statistics given in \eqref{eq:reparam_suff}, 
 we can express the relative log-odds of observing $\bY_{i,j}$ to be $``+"$ abs $``-"$ instead of $``0"$, conditional on $z_i = z_j = k$:
\beno
        \log \left( \dfrac{\mbP_{\bbeta_{w,\operatorname{vec}}}(Y_{i,j} = ``+" \mid\bY_{(-ij)} =\by_{(-ij)}, z_i = z_j = k)}
        {\mbP_{\bbeta_{w,\operatorname{vec}}}(Y_{i,j} = ``0" \mid \bY_{(-ij)} = \by_{(-ij)}, z_i = z_j = k)} \right) &=&\bbeta_{w,\operatorname{vec}}^{\top} \Delta_{i,j,k}^{0 \rightarrow +}(\by) \s\\
        \log \left( \dfrac{\mbP_{\bbeta_{w,\operatorname{vec}}}(Y_{i,j} = ``-" \mid\bY_{(-ij)} = \by_{(-ij)}, z_i = z_j = k)}
        {\mbP_{\bbeta_{w,\operatorname{vec}}}(Y_{i,j} = ``0" \mid \bY_{(-ij)} = \by_{(-ij)}, z_i = z_j = k)} \right) &=&\bbeta_{w,\operatorname{vec}}^{\top} \Delta_{i,j,k}^{0 \rightarrow -}(\by).
\ee
The log-likelihood function is then defined as
\be
\label{eq:llh}
    \ell(\bbeta_{w,\operatorname{vec}}) 
    &=& \dsum_{i < j} \left( \bbeta_{w,\operatorname{vec}}^\top \Delta_{i,j,k}^{0 \rightarrow y_{i,j}}(\by) - \log \dsum_{y} \exp\left(\bbeta_{w,\operatorname{vec}}^\top \Delta_{i,j,k}^{0 \rightarrow y}(\by)\right) \right).
\ee
Newton-Raphson algorithms are commonly used to maximize \eqref{eq:llh} based on \eqref{eq:deriv} with the following gradient and negative Hessian:
\be \label{eq:deriv}
\bu(\bbeta_{w,\operatorname{vec}}) &=& \dfrac{\partial}{\partial \bbeta_{w,\operatorname{vec}}} \ell(\bbeta_{w,\operatorname{vec}}), \quad
\bJ(\bbeta_{w,\operatorname{vec}}) &=& -\dfrac{\partial }{\partial \bbeta_{w,\operatorname{vec}}}\bu(\bbeta_{w,\operatorname{vec}}).
\ee

\subsection{Uncertainty Quantification} \label{sec:uq}

Given the converged block membership probabilities $\hat{\balpha}$, \citet{babkin2020large} assign each node deterministically to the block corresponding to the highest posterior probability, i.e., $\hat{\bz} = (\argmax\limits_{k \, \in\, \{1, \ldots, K\}} \alpha_{i,k})_{i=1}^N$. 
Consecutively, the uncertainty of $\hat{\btheta}$ conditions on this assignment, disregarding any uncertainty from the block allocation. 
Contrasting this approach, we propose to sample $T$ times block assignments from the approximate posterior distributions. 
Given these sampled block memberships, $\bz^{(1)},\ldots,\bz^{(T)}$, the parameters of the model,$(\bbeta_{w,\operatorname{vec}}^{(1)},\bSigma^{(1)}), \ldots, (\bbeta_{w,\operatorname{vec}}^{(T)}, \bSigma^{(T)})$, are re-estimated for each sampled partition to obtain valid uncertainty quantification.
The estimated parameters $\bbeta_{w,\operatorname{vec}}^{(t)}$ are pooled as follows
\beno
\mbE(\bbeta_{w,\operatorname{vec}}) &=& \bar{\bbeta}_{w,\operatorname{vec}}= \dfrac{1}{T} \dsum_{t=1}^T \hat{\bbeta}_{w,\operatorname{vec}}^{(t)}, \s\\
\text{Var}(\bbeta_{w,\operatorname{vec}}) &=& \dfrac{1}{T} \left(\dsum_{t=1}^T \hat{\bSigma}^{(t)} \right) 
+ \dfrac{1}{T-1} \dsum_{t=1}^T \left(\hat{\bbeta}_{w,\operatorname{vec}}^{(t)} - \bar{\bbeta}_{w,\operatorname{vec}}\right)
\left(\hat{\bbeta}_{w,\operatorname{vec}}^{(t)} - \bar{\bbeta}_{w,\operatorname{vec}}\right)^{\top}.
\ee
Standard errors $\hat{\bSigma}^{(t)}$ are computed based on a Taylor approximation derived in Supplement \ref{sec:godambe}.
Making use of the second variant with multiple imputations in Section \ref{sec:conditional_the}, ensures that all these values are already calculated during estimation. 
To assess model fit and enable model selection under this uncertainty quantification, we evaluate the AIC for each sampled partition. By extending the bridge sampler introduced in \citet{hunter2006inference}, we compute the AIC for each sample and report the average across all $T$ partitions. 

\section{Simulation Study} \label{sec:sim}

\begin{figure}[!t]
    \centering
    \includegraphics[width = 0.7\linewidth]{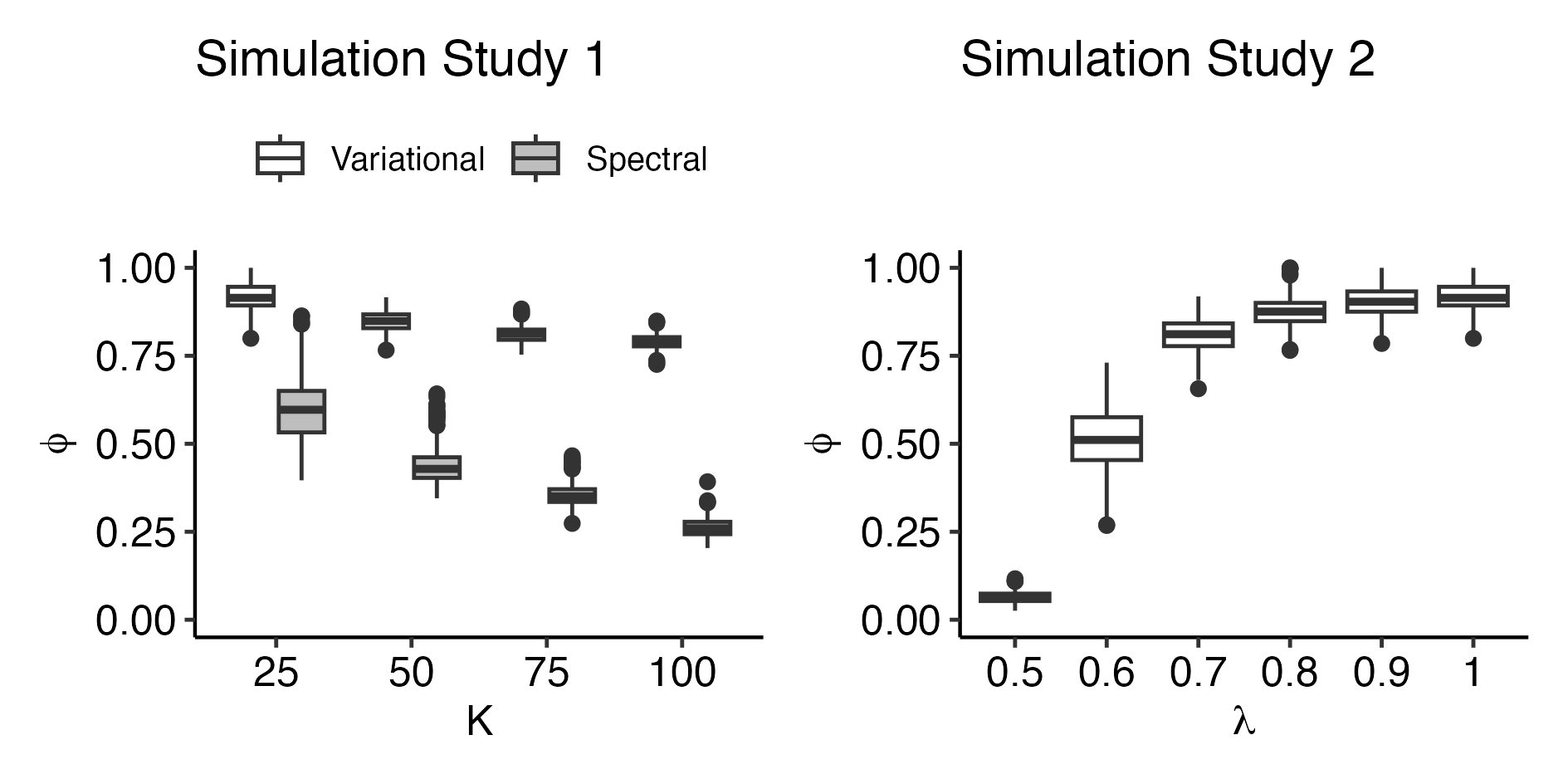}
    \caption{Block recovery performance measured by Yule's $\phi$ coefficient. 
    Left: Simulation Study~1 showing recovery accuracy for different numbers of blocks 
    ($K = 25, 50, 75, 100$) and network sizes ($N = 1{,}250$ to $5{,}000$) at $\lambda = 1$. 
    Right: Simulation Study~2 showing the effect of between-block sparsity, controlled by 
    $\lambda$, on block recovery for $K = 25$. Higher values indicate better agreement with 
    the true block structure.}
    \label{fig:phi_combined}
\end{figure}

\begin{figure}[!t]
    \centering
    \includegraphics[width=\linewidth]{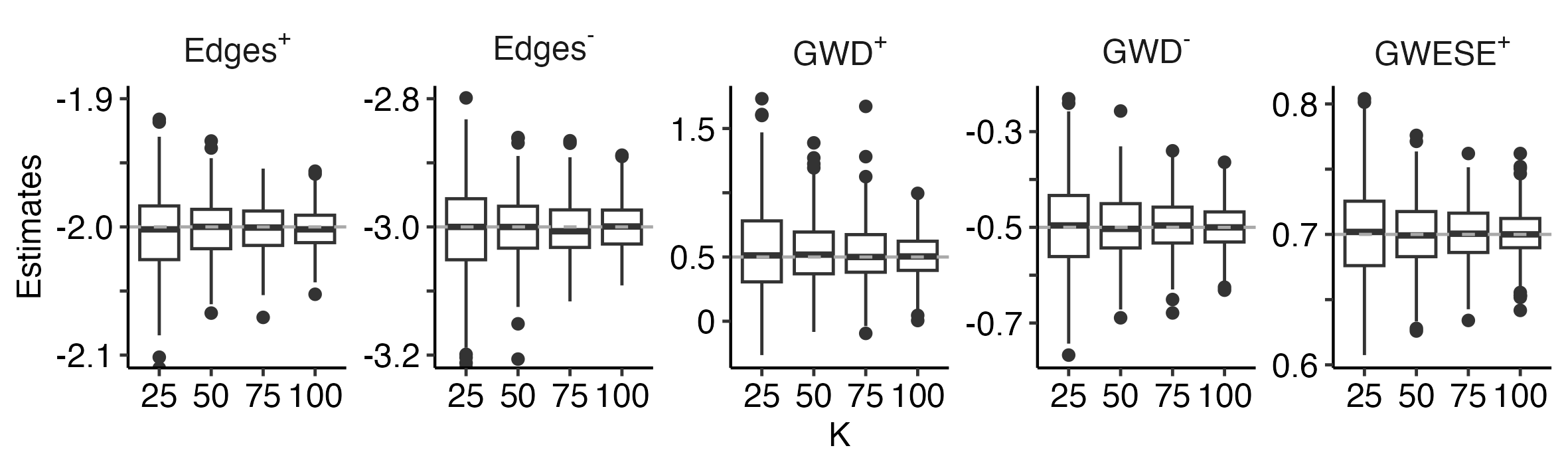}
        \caption{Maximum pseudo-likelihood estimates for within-block parameters in Simulation Study 1. Results are based on known block memberships and used to assess parameter recovery accuracy across increasing network sizes.}
    \label{fig:coef}
\end{figure}

We evaluate the performance of the proposed estimation procedure in both steps. For the evaluation of the parameter recovery in Step 2, we use MPLE as they are more scalable than MCMC-based methods. 
To assess the block recovery, we compare the variational approach described in Supplement \ref{sec:est} to the spectral clustering, for which we treat the network as a binary one. 
We also evaluated spectral clustering using only positive edges and ignoring negative edges, as reported in Section \ref{sec:comp_clust}. 

We conduct two simulation studies to evalute our method's performance. In Simulation Study 1 (Supplement \ref{sec:sim1}), we assess block and parameter recovery across networks of ($N \in \{1250, 2500, 3750, 5000\}$) with $K \in \{25,50,75,100\}$ blocks. Our method consistently outperforms binary spectral clustering in block recovery (Figure \ref{fig:phi_combined}) and accurately estimates model coefficients across all network sizes (Figure \ref{fig:coef}). Robustness to misspecification of $K$ is examined in Supplement \ref{sec:missk}, showing that overestimating $K$ has minimal impact while underestimating gradually reduces accuracy. 

Simulation Study 2 (Supplement \ref{sec:sim2}) examines block recovery under varying sparsity levels by gradually reducing between-block sparsity. As expected, block recovery performance decreases as between-block connectivity increases (lower $\lambda$), as shown in Figure \ref{fig:phi_combined}. Full details of both simulation studies, including parameter specifications and additional results, are provided in the Supplement \ref{sec:sim_supp}.

\section{Wikipedia Network} \label{sec:irl}

We demonstrate our framework in an application to a network of Wikipedia editors contributing to content across a range of topics. 
Wikipedia is a free, community-edited online encyclopedia that has been the subject of numerous prior studies \citep{brandes2009network,iba2010analyzing}. 
A subset of these editors can be considered experts in specific fields, as they focus their contributions to a limited number of Wikipedia pages within their area of expertise.
These editors typically exhibit limited awareness of edits or interactions occurring outside their domain. 
Such behavior gives rise to block structure and localized dependence within the network, where each block corresponds to a distinct area of knowledge.

The edges in the network were constructed as follows: a negative edge between users $i$ and $j$ is added if either user undoes or deletes the other's work, and a positive edge if either user redoes the other's work. 
The raw dataset includes data from more than \mbox{10,000} articles and is described in detail in \citet{lerner2019network}.
To extract our subsample of experts, we filtered the data from the full network of \mbox{1,634,189} editors.
First, to ensure that we focus on meaningful contributions, we included only users who added at least 100 words per page. 
We, thereby, distinguish between users who primarily add substantive new content, likely indicating expertise on specific topics, from those who make frequent but less substantive edits, such as corrections or anti-vandalism activities. 
Second, we excluded users who contributed to more than 10 pages, as this typically indicates bots or users focused on maintenance rather than topic-specific contributions. 
From the remaining users, we selected 50 pages whose names are provided in Table \ref{sec:app_wiki}. 
This procedure resulted in an undirected network of $N = \mbox{2,115} $ nodes.

\subsection{Model Specification} \label{sec:oos}

We apply the two-stage estimation approach to this network. 
For this, we assume that each page is an individual block, resulting in 50 blocks in the network. Note, however, that there is no ground truth block membership for the nodes, only for the edges, so it is possible that the number of blocks is less than 50 in the case where two pages contain a similar subset of users. Conversely, the number of blocks could be higher than 50 if a single page includes heterogeneous sections that correspond to distinct user groups. That being said, the Yule's phi coefficient between the estimated block membership and the page an editor contributed to most is 0.69, indicating a moderately strong alignment between the assumed and inferred structures. To account for potential misclassifications in the inferred structure, we apply an uncertainty correction to adjust for estimation error in block assignments.

\begin{figure}[!t]
    \centering
    \includegraphics[width=\linewidth]{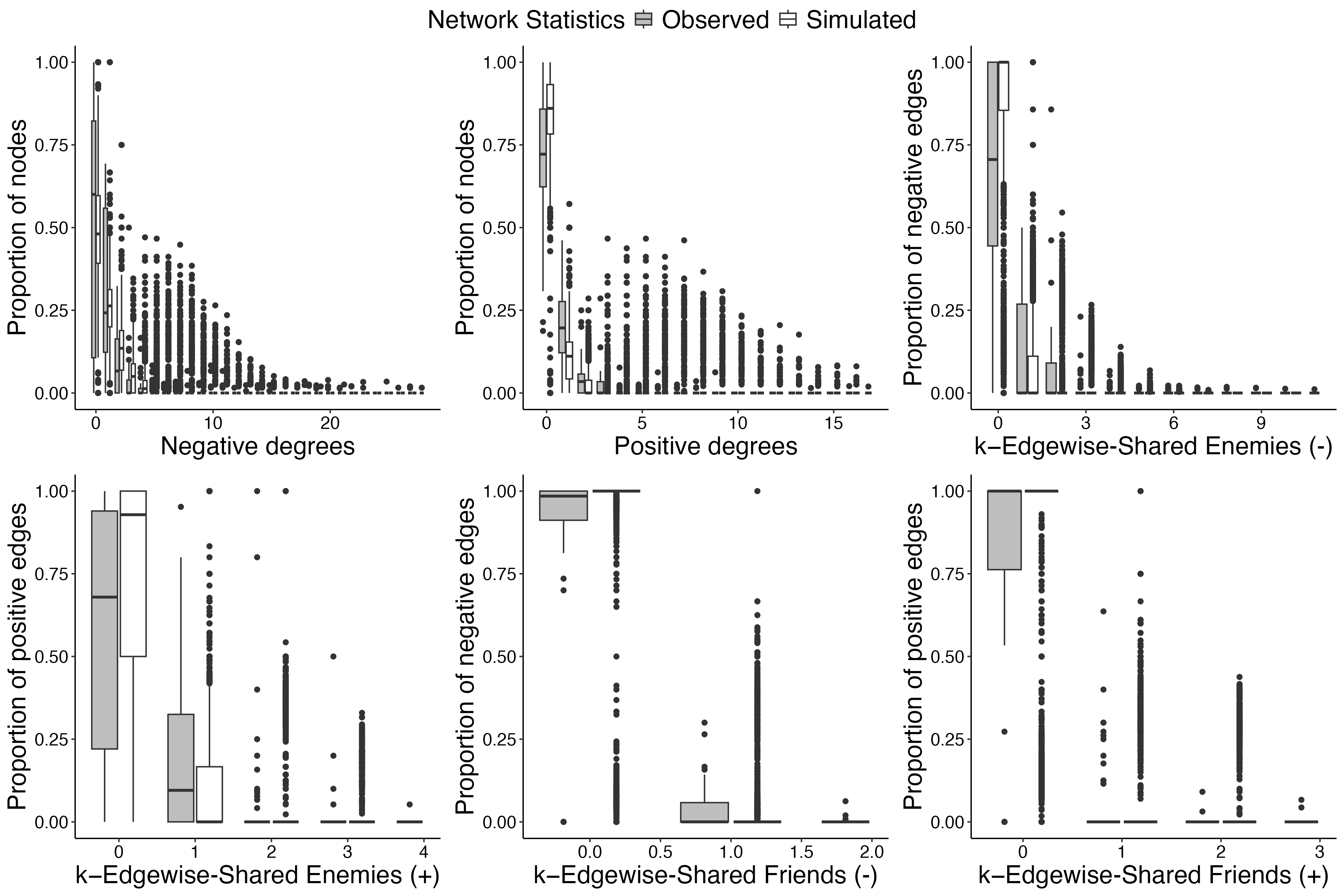}
        \caption{Out-of-sample cross-validation results for the ``Full Triad" model including all triadic terms ($\text{Edges}^{+/-}$, $\text{GWD}^{+/-}$, and all GWESP). The distribution of simulated statistics across 100 replications is compared against the observed statistics for each block.}
    \label{fig:full_oos}
\end{figure}

To decide which model specifications are best suited to describe the observed Wikipedia network, we adapt the out-of-sample cross-validation for multi-level networks used by \citet{stewart2019multilevel} to our setting. 
We remove one block and estimate a model for the remaining 49 blocks. The resulting model coefficients are used to individually simulate the previously excluded block 100 times. 
To best capture signed network structure, we  plot the distribution of the observed network statistics against the simulated statistics: degree distributions (positive and negative), edgewise shared friends ($\text{ESF}^+$ and $\text{ESF}^-$), and edgewise shared enemies ($\text{ESE}^+$ and $\text{ESE}^-$). 
In total, we tested four model specifications of increasing complexity, which can be seen in Table \ref{tbl:ergm_within_between}. 
The decay term for the geometrically weighted degrees and for the geometrically weighted shared partners (GWESP) it is set to $.2$.
The vector of sufficient statistic for the ``Full Triad'' model is thus given by:
\beno 
\bs(\by_{k,k}) &=& (&\hspace*{-10pt}
 \text{Edges}^+(\by_{k,k}), \,
 \text{Edges}^-(\by_{k,k}), \,
 \text{GWD}^+(\by_{k,k}, .2), \,
 \text{GWD}^-(\by_{k,k}, .2), \\
 &~ &~ & \hspace*{-10pt}\text{GWESE}^+(\by_{k,k}, .2), \,
 \text{GWESF}^+(\by_{k,k}, .2), \,
 \text{GWESE}^-(\by_{k,k}, .2), \,
 \\
 &~ &~ & \hspace*{-10pt} \text{GWESF}^-(\by_{k,k}, .2) )^\top.
 \ee
The results of the out-of-sample cross-validation, shown in Figure \ref{fig:full_oos} and Supplement \ref{sec:oos_sup}, suggest that the ``Full Triad" model, including edges, GWD and all GWESP terms, is better equipped to describe the Wikipedia network.

\subsection{Results} \label{sec:res}

\begin{table}[!t] \label{tab:res}
\caption{Estimated within-block parameters with uncertainty correction ($T = 100$). Coef. indicates the estimated parameter, and SE is the associated standard error, based on the estimated Godambe information matrix, using 100 simulations per block. Models compared are: I (Independent), I+D (Degree), I+D+PT (Partial Triad), and I+D+FT (Full Triad). $\Delta$AIC values indicate the difference in AIC relative to the independent model.\label{tbl:ergm_within_between}}
\centering
\begin{tabular}{l rr c rr c rr c rr}

 & \multicolumn{2}{c}{Independent} & & \multicolumn{2}{c}{Degree} & & \multicolumn{2}{c}{Partial Triad} & & \multicolumn{2}{c}{Full Triad} \\
\cmidrule(lr){2-3} \cmidrule(lr){5-6} \cmidrule(lr){8-9} \cmidrule(lr){11-12}
Parameters      & Coef. & SE & & Coef. & SE & & Coef. & SE & & Coef. & SE \\
\hline
$\text{Edges}^+$              & $.581$ & $.402$ & & $2.174$ & $.266$ & & $.849$ & $.469$ & & $.300$ & $.619$ \\
\quad$\times\log(N_k)$ & $-1.130$ & $.087$ & & $-1.072$ & $.053$ & & $-.871$ & $.087$ & & $-.822$ & $.110$ \\
$\text{Edges}^-$              & $.672$ & $.265$ & & $2.022$ & $.179$ & & $1.510$ & $.205$ & & $.691$ & $.278$ \\
\quad$\times\log(N_k)$ & $-.946$ & $.057$ & & $-1.053$ & $.037$ & & $-.974$ & $.041$ & & $-.860$ & $.052$ \\
$\text{GWD}^+$                &         &         & & $-1.475$ & $.089$ & & $-1.531$ & $.131$ & & $-1.364$ & $.161$ \\
$\text{GWD}^-$                &         &         & & $-.985$ & $.069$ & & $-.987$ & $.070$ & & $-.819$ & $.080$ \\
 $\text{GWESE}^+$              &         &         & &         &         & & $1.584$ & $.161$ & & $1.599$ & $.208$ \\
$\text{GWESF}^+$              &         &         & &         &         & &         &         & & $.095$ & $.271$ \\
$\text{GWESE}^-$              &         &         & &         &         & &         &         & & $.187$ & $.074$ \\
$\text{GWESF}^-$              &         &         & &         &         & &         &         & & $.808$ & $.372$ \\
\hline
$\Delta$AIC & \multicolumn{2}{c}{$0$} & & \multicolumn{2}{c}{$951$} & & \multicolumn{2}{c}{$\mbox{1,360}$} & & \multicolumn{2}{c}{$\mbox{1,382}$} \\
\hline
\end{tabular}%
\end{table}

As described in Section \ref{sec:est}, we sample 100 times from the block membership probability to account for the uncertainty within latent block memberships. 
The results of the within-block estimation are provided in Table \ref{tbl:ergm_within_between}. 
Standard errors were also approximated based on 100 simulations. 
The interpretation of these coefficients follows the standard SERGM logic introduced in Section \ref{sec:exis}. The only difference is that all probabilities and change statistics are interpreted conditional on the block membership vector $\bz$. Because blocks are independent of one another, the coefficients are directly comparable across blocks. Differences in observed edge probabilities arise from block size and local network structure rather than from block-specific parameters.

The edges terms are modeled as a function of block size $\log(N_k)$. 
The terms $\text{Edges}^+$ and $\text{Edges}^-$ represent the intercepts, which alone are not directly interpretable, as they would only apply to hypothetical blocks of size 1. 
Instead, the combination of intercept and block size effect reflects the expected edge probability within a given block. 
Across all models, the coefficients for positive edges are lower than for negative edges, indicating that negative edges are more frequent within blocks. 
The consistently negative coefficients for the block size terms demonstrates that larger blocks tend to be sparser, with edge probability decreasing as block size increases. 
This reflects the fact that the number of possible edges grows quadratically, while the number of actual edges grows at a slower pace. 
Adding GWD terms reduces the edge coefficients, suggesting that some variation in edge probability is explained by degree structure. 
The GWD coefficients are negative for both edge types in all models, indicating a general tendency against high-degree nodes. 
The effect is stronger for positive edges, implying that these edges are more evenly distributed and less likely to form hubs.
As expected from structural balance theory, the $\text{GWESE}^+$ terms have positive and statistically significant coefficients in both the ``Full Triad" and ``Partial Triad" model. This means that users are more likely to restore the work of another user if they share a common enemy.
While still positive, the other balanced triad represented by $\text{GWESF}^+$ is closer to zero, as is the unbalanced triad  $\text{GWESE}^-$. 
However, the term $\text{GWESF}^-$ is significantly larger than $\text{GWESF}^+$, which is not consistent with structural balance theory. 
This discrepancy suggests that structural balance theory may not fully apply to this expert network, For instance, undoing a friend-of-a-friend's contribution could be more common due to task-related disagreements rather than group-based antagonism.
Overall, the more complex models capture important network features, such as triadic relationships, and this is reflected in their lower AIC values, indicating better model fit despite increased complexity.

Our Wikipedia network analysis of thousands of subject-matter experts is one example where global dependence assumption is unrealistic. 
Through the use of local dependence,  we found some adherence to the behavioral norms predicted by structural balance theory. 
For instance, users with a common enemy are more likely to restore each other's contributions, which is consistent with \citet{lerner2019network}.
To validate the results in Table \ref{tbl:ergm_within_between}, we conducted in Supplement \ref{sec:gof_sup} a goodness-of-fit analysis following \citet{hunter2008goodness}.

\section{Conclusion} \label{sec:con}

Our proposed model for signed networks combines the strengths of stochastic block models and exponential random graph models, while being scalable to thousands of nodes. 
Several directions for extending the proposed model exist: 
First, we assumed that the number of blocks, $K$, is known, which may not hold in practice. 
The choice of $K$ remains an open issue in the SBM literature.
Methods to infer $K$ from data would enhance model flexibility \citep{saldana2017many}. 
Second, the assumption of non-overlapping community structure could be relaxed by allowing nodes to belong to multiple blocks, as in mixed-membership models \citep{latouche2011}.
This would accommodate more realistic block structures.
Third, the observed degree heterogeneity in the Wikipedia network highlights the need for a degree-corrected stochastic block model for signed networks. 

\section*{Acknowledgments}
The authors are indebted to the constructive comments and suggestions by Michael Schweinberger, which have led to numerous improvements. 
The authors would also like to thank the HPC Service of FUB-IT, Freie Universität Berlin, for computing time \citep{bennett2020curta}.

\bibliographystyle{chicago}
\bibliography{references}

\newpage

\begin{center}
{\LARGE\textbf{Supplementary Materials:\\ 
Scalable Signed Exponential Random Graph Models under Local Dependence}}
\end{center}

\setcounter{section}{0}

\setcounter{page}{1}

\appendix
\numberwithin{equation}{section}
\renewcommand{\cftsecfont}{\mdseries}
\renewcommand{\cftsecpagefont}{\mdseries}
\setlength{\cftbeforesecskip}{0 pt} 
\renewcommand{\cftsecleader}{\cftdotfill{\cftdotsep}} 
\startcontents
\printcontents{ }{1}{}

\newpage

\section{Uncertainty Quantification with known Blocks} \label{sec:godambe}

When using MPLE, the standard errors $\hat{\bSigma}^{(t)}$ obtained from a logistic regression are, as \citesupp{strauss1990pseudolikelihood} and \citesupp{van2009framework} pointed out, inappropriate because dyadic independence is assumed.
Similar to \citesupp{schmid2023computing}, we approximate the variance of the MPLE by a first-order Taylor approximation of the pseudo-score function $\bu(\btheta)$ in some $\btheta$ evaluated at the MPLE, denoted by $\hat{\btheta}$. 
By definition, we then get $\bu(\hat{\btheta}) = 0$, which yields
\be
\label{eq:taylor}
0 &\approx& \bu(\btheta)  - \bJ(\btheta) \, (\hat{\btheta} - \btheta),  
\ee
where $\bJ(\btheta) = -\dfrac{\partial }{\partial \btheta \bu(\btheta)} \bu(\btheta)$ defines the negative pseudo-Hessian. 
Upon isolating $\hat{\btheta} - \btheta$, the variance of $\hat{\btheta}$ is approximately
\beno
\text{Var} (\hat{\btheta}) &\approx &\text{Var}\left(\bJ(\btheta)^{-1}\,\bu(\btheta)\right). 
\ee
Plugging in the converged MPLE $\hat{\btheta}$ for $\btheta$ yields the following variance estimator: 
\be
\label{eq:taylor}
\text{Var} (\hat{\btheta}) &\approx& \text{Var}\left(\bJ(\hat{\btheta})^{-1}\,\bu(\hat{\btheta})\right). 
\ee
The $\text{Var}$ operator is taken with respect to $\bY \mid \bZ$ following  the distribution defined in Equation \ref{eq:decomp} characterized by $\hat{\btheta}$. 
Both terms, the pseudo score and negative Hessian, are random functions of $\bY \mid \bZ$:
\beno
p_{i,j}^y &=& \mbP(Y_{i,j} = y \mid \bY_{(-ij)} = \by_{(-ij)}) = 
    \dfrac{\exp\left(\btheta^\top\Delta_{i,j}^{0 \rightarrow y}\right)}
          {\dsum_{t \in \mathcal{S}} \exp\left(\btheta^\top\Delta_{i,j}^{0 \rightarrow t}\right)} \s\\

\bu(\btheta) &=& \dfrac{\partial}{\partial \btheta} \ell_{PL}(\btheta; \bY, \bZ) = 
\dsum_{i<j} \left(\Delta_{i,j}^{0 \rightarrow y_{i,j}}- \dsum_{y \in \mathcal{S}} p_{i,j}^y\Delta_{i,j}^{0 \rightarrow y} \right) \s\\

\bJ(\btheta) &=& -\dfrac{\partial^2}{\partial \btheta \partial \btheta^T} \ell_{PL}(\btheta; \bY, \bZ) \s\\
&=&  \dsum_{i<j} \left(\dsum_{y}p_{i,j}^y\Delta_{i,j}^{0 \rightarrow y}(\Delta_{i,j}^{0 \rightarrow y})^\top
                    - \left(\dsum_y p_{i,j}^y\Delta_{i,j}^{0 \rightarrow y}\right)
                    \left(\dsum_{s}p_{i,j}^s\Delta_{i,j}^{0 \rightarrow y}\right)^\top
                        \right).
\ee
Therefore, we simulate $R$ networks using MCMC algorithms and 
calculate for each sample the inverse negative pseudo-Hessian and pseudo-score function. 
For the $r$th sample, both terms are denoted by $\bu^{(r)}(\hat{\btheta})$ and $\bJ^{(r)}(\hat{\btheta})^{-1}$ with $\bv^{(r)}(\hat{\btheta}) \coloneqq \bJ^{(r)}(\hat{\btheta})^{-1}\,\bu^{(r)}(\hat{\btheta})$. 
We then approximate \eqref{eq:taylor} by
\beno
    \text{Var} (\hat{\btheta}) &\approx& \dfrac{1}{R-1} \dsum_{r=1}^R 
                                    \left(\bv^{(r)}(\hat{\btheta}) - \bar{v}(\hat{\btheta})\right) 
                                    \left(\bv^{(r)}(\hat{\btheta}) - \bar{v}(\hat{\btheta})\right),
\ee
where $\bar{v}(\hat{\btheta}) \coloneqq 1/R \, \sum_{r=1}^R \bv^{(r)}(\hat{\btheta})$ denotes the sample mean of the score vectors.

\hide{
standard errors by calculating the Godambe matrix \citesupp{godambe1960optimum}. The Godambe matrix is 
\beno
\bG(\btheta^{(t)}) = \bJ(\btheta^{(t)})^{-1}\bV(\btheta^{(t)})\bJ(\btheta^{(t)})^{-1},
\ee
where $\bJ(\btheta^{(t)})$ is defined in \eqref{eq:deriv} and $\bV(\btheta^{(t)}) \coloneqq \text{Var}(\bu(\btheta^{(t)}))$ is called the variability matrix. 
However, $\bV(\btheta^{(t)})$ cannot be directly computed and must be approximated by 

}

\section{Computation} \label{sec:comp}

\subsection{Surrogate Function} \label{sec:surrogate}
In order to detail how the updates of $\balpha$ are carried out using sparse matrix operations, we introduce a surrogate function $Q({\bgamma}^{(t)},{\btheta}^{(t)};{\balpha}^{(t)},{\balpha})$ that provides a tractable lower bound that can be maximized iteratively and rearrange it as follows:
\beno
    Q({\bgamma}^{(t)},{\btheta}^{(t)};{\balpha}^{(t)},{\balpha})  &=&\dsum_{i<j}\dsum_{k=1}^K\dsum_{l=1}^K
    \left(\alpha_{i,k}^2\dfrac{\alpha_{j,l}^{(t)}}{2\alpha_{i,k}^{(t)}}+\alpha_{j,l}^2\dfrac{\alpha_{i,k}^{(t)}}{2\alpha_{j,l}^{(t)}}\right)
    \log p_{k,l}(y_{i,j}) \s\\
    &+&\dsum_{i=1}^N\dsum_{k=1}^K\alpha_{i,k}\left(\log\gamma_k^{(t)}-\log\alpha_{i,k}^{(t)}-\dfrac{\alpha_{i,k}}{\alpha_{i,k}^{(t)}}+1\right) \s\\
    &=& \dsum_{i=1}^N \dsum_{k=1}^K \dfrac{\Omega_{i,k}^{(t)}(\by,\balpha^{(t)},\btheta^{(t)})}{\alpha_{i,k}^{(t)}}\alpha_{i,k}^{(t)}\s\\
    &+&\dsum_{i=1}^N\dsum_{k=1}^K\alpha_{i,k}\left(\log\gamma_k^{(t)}-\log\alpha_{i,k}^{(t)}-\dfrac{\alpha_{i,k}}{\alpha_{i,k}^{(t)}}+1\right) \s\\
    &=& \dsum_{i=1}^N \dsum_{k=1}^K \left(\dfrac{\Omega_{i,k}^{(t)}(\by,\balpha^{(t)},\btheta^{(t)})}{\alpha_{i,k}^{(t)}} - \dfrac{1}{\alpha_{i,k}^{(t)}}\right)\alpha_{i,k}^2\s\\
    &+& \dsum_{i=1}^N \dsum_{k=1}^K \left(\log\gamma_k^{(t)}-\log\alpha_{i,k}^{(t)} +1\right)\alpha_{i,k} \s\\
    &=& \dsum_{i=1}^N \dsum_{k=1}^K A_{i,k}(\by,\balpha^{(t)})\alpha_{i,k}^2 + 
    B_{i,k}(\bgamma^{(t)}, \balpha^{(t)})\alpha_{i,k}
\ee
with
\be
\label{eq:quadratric}
    \Omega_{i,k}^{(t)}(\by,\balpha^{(t)},\btheta^{(t)}) &\coloneqq& \dsum_{j\neq i}^N\sum_{l=1}^K \alpha_{j,l}^{(t)}\log p_{k,l}(y_{i,j}),
\ee
the quadratic term 
\be
\label{eq:linear}
    A_{i,k}(\by,\balpha^{(t)}) \coloneqq \dfrac{\Omega_{i,k}^{(t)}(\by,\balpha^{(t)},\btheta^{(t)})}{ \alpha_{i,k}^{(t)}} - \dfrac{1}{\alpha_{i,k}^{(t)}}
\ee
and the linear term of the quadratic problem
\beno
     B_{i,k}(\bgamma^{(t)}, \balpha^{(t)}) \coloneqq \log\gamma_k^{(t)}-\log\alpha_{i,k}^{(t)} +1.
\ee

\subsection{Sparse Matrix Multiplication} \label{sec:sparse}
To update our estimate of $\balpha$, we need to evaluate $A_{i,k}(\by, \balpha^{(t)})$ and $B_{i,k}(\bgamma^{(t)}, \balpha^{(t)})$, with the former being the problematic part, giving us a complexity of $\mathcal{O}(N^2K^2)$. Therefore, we decompose $\Omega_{i,k}^{(t)}(\by, \balpha^{(t)}, \btheta^{(t)})$ into two parts: one when the network is completely empty, i.e., $y_{i,j} = 0$ for all $i,j = 1, \ldots, N$, and the other capturing all dyads where $y_{i,j} \ne 0$, which allows us to exploit the sparse nature of the network:
\be
    \Omega_{i,k}^{(t)}(\by,\balpha^{(t)},\btheta^{(t)}) &\coloneqq& \dsum_{j\neq i}^N\sum_{l=1}^K \alpha_{j,l}^{(t)}\log p_{k,l}^{(t)}(y_{i,j}) \s\\
    &=& \dsum_{j\neq i}^N\sum_{l=1}^K \alpha_{j,l}^{(t)}\log p_{k,l}^{(t)} (``0") +\dsum_{j\neq i}^N\mI(y_{i,j}= ``+")\sum_{l=1}^K \alpha_{j,l}^{(t)}
    \log\dfrac{ p_{k,l}^{(t)}(``+")}{ p_{k,l}^{(t)} (``0")} \s\\
    &+&\dsum_{j\neq i}^N\mI(y_{i,j}= ``-")\sum_{l=1}^K \alpha_{j,l}^{(t)}
    \log\dfrac{ p_{k,l}^{(t)}(``-")}{ p_{k,l}^{(t)} (``0")} \s\\
    &=& \Omega_{i,k}^{(t)}(\bm{0},\balpha^{(t)}) + \Lambda_{i,k}^{(t)}(\by,\balpha^{(t)}) \label{eq:omega}
\ee
with
\beno
    \Lambda_{i,k}^{(t)}(\by,\balpha^{(t)}) &=& \dsum_{j\neq i}^N\mI(y_{i,j}= ``+")\sum_{l=1}^K \alpha_{j,l}^{(t)}
    \log\dfrac{ p_{k,l}^{(t)}(``+")}{ p_{k,l}^{(t)} (``0")} \s\\
    &+&\dsum_{j\neq i}^N\mI(y_{i,j}= ``-")\sum_{l=1}^K \alpha_{j,l}^{(t)}
    \log\dfrac{ p_{k,l}^{(t)}(``-")}{ p_{k,l}^{(t)} (``0")}
\ee

The first part, in which we assume the network is completely empty, can be computed using matrix multiplication as follows:
\beno
    \Omega_{i,k}^{(t)}(\bm{0},\balpha^{(t})&=\dsum_{j \neq i}^N\dsum_{l=1}^K\alpha_{j,l}^{(t)}\log p_{k,l}^{(t)} (``0") \s\\
    &=\dsum_{l=1}^K\bigg(\underbrace{\dsum_{j=1}^N\alpha_{j,l}^{(t)}}_{\coloneqq \tau_l^{(t)}}-\alpha_{i,l}^{(t)}\bigg)\log p_{k,l}^{(t)} (``0") \s\\
    &=\dsum_{l=1}^K\left(\tau_l^{(t)}-\alpha_{i,l}^{(t)}\right)\log p_{k,l}^{(t)} (``0")
\ee
Define 
\beno
\bA_{\bm{0}}^{(t)}&\coloneqq&\left(\begin{array}{cccc}
\tau_1^{(t)}-\alpha_{1,1}^{(t)} & \tau_2^{(t)}-\alpha_{1,2}^{(t)} & \ldots & \tau_K^{(t)}-\alpha_{1, K}^{(t)} \\
\tau_1^{(t)}-\alpha_{2,1}^{(t)} & \tau_2^{(t)}-\alpha_{2,2}^{(t)} & \ldots & \tau_K^{(t)}-\alpha_{2, K}^{(t)} \\
\vdots & \vdots & \ddots & \vdots \\
\tau_1^{(t)}-\alpha_{N ,1}^{(t)} & \tau_2^{(t)}-\alpha_{N, 2}^{(t)} & \ldots & \tau_K^{(t)}-\alpha_{N, K}^{(t)}
\end{array}\right)
\ee
and
\beno
\bm{P}_{\bm{0}}^{(t)}&\coloneqq&\left(\begin{array}{cccc}
\log  p_{1,1}^{(t)} (``0") & \log  p_{1,2}^{(t)} (``0") & \ldots & \log  p_{1 ,K}^{(t)} (``0")\\
\log  p_{2,1}^{(t)} (``0") & \log  p_{2,2}^{(t)} (``0") & \ldots & \log  p_{2 ,K}^{(t)} (``0") \\
\vdots & \vdots & \ddots & \vdots \\
\log  p_{K ,1}^{(t)} (``0") & \log  p_{K ,2}^{(t)} (``0") & \ldots & \log  p_{K, K}^{(t)} (``0")
\end{array}\right).
\ee
Then $\Omega_{i,k}^{(t)}(\bm{0},\balpha^{(t)}, \btheta ^{(t)})$ is given by the $(i, k)$ entry of $\bA_{\bm{0}}^{(t)}\bm{P}_{\bm{0}}^{(t)}$.

Next, we correct the error arising from the assumption of an entirely empty network by calculating the term where this assumption does not apply:
\beno
    \Lambda_{i,k}^{(t)}(\by,\balpha^{(t)}) &=& \dsum_{j\neq i}^N\mI(y_{i,j}= ``+")\sum_{l=1}^K \alpha_{j,l}^{(t)}
    \log\dfrac{ p_{k,l}^{(t)}(``+")}{ p_{k,l}^{(t)} (``0")} \s\\
    &+&\dsum_{j\neq i}^N\mI(y_{i,j}= ``-")\sum_{l=1}^K \alpha_{j,l}^{(t)}
    \log\dfrac{ p_{k,l}^{(t)}(``-")}{ p_{k,l}^{(t)} (``0")}
\ee
Define
\beno
\bm{P}_{\bm{+}}^{(t)}&\coloneqq&\left(\begin{array}{cccc}
\log \dfrac{ p_{1,1}^{(t)} (``+")}{ p_{1,1}^{(t)} (``0")} & \log \dfrac{ p_{1,2}^{(t)} (``+")}{ p_{1,2} ^{(t)}(``0")} & \ldots & \log \dfrac{ p_{1,K} ^{(t)}(``+")}{ p_{1,K} ^{(t)}(``0")} \\
\log \dfrac{ p_{2,1} ^{(t)}(``+")}{ p_{2,1} ^{(t)}(``0")} & \log \dfrac{ p_{2,2}^{(t)} (``+")}{ p_{2,2} ^{(t)}(``0")} & \ldots & \log \dfrac{ p_{2,K} ^{(t)}(``+")}{ p_{2,K} ^{(t)}(``0")} \\
\vdots & \vdots & \ddots & \vdots \\
\log \dfrac{ p_{K,1}^{(t)} (``+")}{ p_{K,1} ^{(t)}(``0")} & \log \dfrac{ p_{K,2}^{(t)} (``+")}{ p_{K,2}^{(t)} (``0")} & \ldots & \log \dfrac{ p_{K,K}^{(t)} (``+")}{ p_{K,K}^{(t)} (``0")}
\end{array}\right)
\ee
and
\beno
\bm{P}_{\bm{-}}^{(t)} &\coloneqq&\left(\begin{array}{cccc}
\log \dfrac{ p_{1,1}^{(t)} (``-")}{ p_{1,1}^{(t)} (``0")} & \log \dfrac{ p_{1,2} ^{(t)}(``-")}{ p_{1,2}^{(t)} (``0")} & \ldots & \log \dfrac{ p_{1,K} ^{(t)}(``-")}{ p_{1,K} ^{(t)}(``0")} \\
\log \dfrac{ p_{2,1} ^{(t)}(``-")}{ p_{2,1}^{(t)} (``0")} & \log \dfrac{ p_{2,2} ^{(t)}(``-")}{ p_{2,2}^{(t)} (``0")} & \ldots & \log \dfrac{ p_{2,K}^{(t)} (``-")}{ p_{2,K} ^{(t)}(``0")} \\
\vdots & \vdots & \ddots & \vdots \\
\log \dfrac{ p_{K,1} ^{(t)}(``-")}{ p_{K,1}^{(t)} (``0")} & \log \dfrac{ p_{K,2} ^{(t)}(``-")}{ p_{K,2}^{(t)} (``0")} & \ldots & \log \dfrac{ p_{K,K} ^{(t)}(``-")}{p_{K,K} ^{(t)}(``0")}
\end{array}\right).
\ee
Then $\Lambda_{i,k}^{(t)}(\by,\balpha^{(t)})$ is given by the $(i, k)$ entry of $\by_+\balpha^{(t)}\bm{P}_+^{(t)} + \by_-\balpha^{(t)}\bm{P}_-^{(t)}$. 

Going back to \eqref{eq:omega}, we have to evaluate for each update $\bA_{\bm{0}}^{(t)}\bm{P}_{\bm{0}}^{(t)}$ and $\by_+\balpha^{(t)}\bm{P}_+^{(t)} + \by_-\balpha^{(t)}\bm{P}_-^{(t)}$ yielding in the respective $(i, k)$th entry $\Omega_{i,k}^{(t)}(\bm{0},\balpha^{(t)}, \btheta ^{(t)})$ and $\Lambda_{i,k}^{(t)}(\by,\balpha^{(t)})$. 
This breaks the bottleneck in \eqref{eq:quadratric} of evaluating $\Omega_{i,k}^{(t)}(\by,\balpha^{(t)},\btheta^{(t)})$  
for $i = 1, ..., N$ and $k = 1, ..., K$. 

\subsection{Update Rules} \label{sec:update}
In this section we provide details on the update rules of the estimated block membership probability $\balpha \in [0,1]^{N \times K}$ for each node $n$ for each block $k$, the prior block probability $\bgamma_{k} \in [0,1]$ for each block $k$ and the edge probability $p_{k,l}\in [0,1]$ between each pair of blocks $(k,l)$. These parameters are required for the first step of our two step estimation approach.
The updated rules of $\balpha$, $\gamma_k$ and $p_{k,l}(y_{i,j})$ follow
\beno
    \balpha^{(t+1)} &\coloneqq& \underset{\balpha}{\argmax} \,Q(\bgamma^{(t)},\btheta^{(t)},\balpha^{(t)};\balpha) \s\\
    \gamma_{k}^{(t+1)} &\coloneqq& \dfrac{1}{N} \dsum_{i=1}^N \alpha_{i,k}^{(t+1)}, \, \text{ for } k = 1,\dots,K \s\\
    p_{k,l}^{(t+1)}(y)
    &\coloneqq&
    \dfrac{\dsum_{i < j } \alpha_{i,k}^{(t+1)} \alpha_{j,l}^{(t+1)} \mI(y_{i,j}=y)}{\dsum_{i < j }\alpha_{i,k}^{(t+1)} \alpha_{j,l}^{(t+1)}} 
\ee
for $k,l = 1,\dots,K$ and $y\in  \mathcal{S} \coloneqq \{\!``-",``0",``+"\!\}$.

\section{Simulation Study} \label{sec:sim_supp}

We assess block recovery using the Yule's block labeling-invariant $\phi$-coefficient:
\beno 
\Phi(\bz^{\ast}, \bz) &=& \dfrac{n_{0,0} n_{1,1} - n_{0,1} n_{1,0}}{\sqrt{(n_{0,0} + n_{0,1})(n_{1,0} + n_{1,1})(n_{0,0} + n_{1,0})(n_{0,1} + n_{1,1})}},
\ee
where $n_{0,0}$ represents the number of node pairs assigned to different blocks, while $n_{1,1}$ counts the pairs assigned to the same block in both. 
The sum $n_{0,1} + n_{1,0}$ counts how often the assignments differ. 
The coefficient of 1 demonstrates full aggreement between $\bz^\star$ and $\bz$.

\subsection{Simulation Study 1: Block \& Parameter Recovery.} \label{sec:sim1}
We simulate networks using the model specification detailed in Example~3  in Section \ref{sec:model} with geometrically weighted degrees ($\text{GWD}^{+/-}$) and edgewise shared enemies ($\text{GWESE}^+$) statistics. 
The number of blocks $K$ each of size 50 on a grid, $K \in \{25, \,50, \,75, \,100\}$, resulting in networks of size $N \in \{\mbox{1,250},\,  \mbox{2,500}, \,\mbox{3,750}, \,\mbox{5,000}\}$.
The within-block coefficients are set to  $\btheta_{k,k} = ( -2,0.5, -3, -0.5, 0.7 )^\top$
corresponding to the sufficient statistics $\text{Edges}^+$, $\text{GWD}^+$, $\text{Edges}^-$, $\text{GWD}^-$, and $\text{GWESE}^+$, respectively.
The between-block coefficients, corresponding to positive and negative edges, are set to $\btheta_{k,l} = (-1.5, -0.5)^\top \log(N)$. 
The log-scaling with $N$ allows for sparsity in the between-block networks as the total number of nodes increases.

These parameter values are chosen to ensure that the simulated networks meet two criteria. 
First, the networks exhibit a clear block structure with denser connectivity within blocks than between blocks.
Second, the networks resemble real-world signed social networks: the positive $\text{GWD}^+$ coefficient encourages the formation of nodes with high positive degree, inducing centralization, while the negative $\text{GWD}^-$ coefficient suppresses high negative degree nodes. 
The positive $\text{GWESE}^+$ coefficient reflects structural balance theory, favoring balanced triads where the \textsl{``enemy of my enemy is my friend"}.

The generated networks are used to test block recovery by comparing them to binary spectral clustering (see Figure \ref{fig:phi_combined}) and to assess parameter recovery using MPLE when true block membership is known (see Figure \ref{fig:coef}). 
Our estimation method consistently outperforms spectral clustering in recovering block structure and accurately estimates model coefficients across different numbers of blocks.

\subsection{Simulation Study 2: Sparsity.} \label{sec:sim2}
We conduct a simulation study to evaluate block recovery under varying levels of sparsity. 
The considered networks consist of $K = 25$ blocks, each containing 50 nodes, resulting in $N = \mbox{1,250}$ nodes in total. 
The within-block coefficients remain the same as in the previous scenario. 
To assess how block recovery is influenced by between-block sparsity, we set the coefficients for positive and negative between-block edges to 
$\btheta_{k,k} = (-1.5, -0.5)^\top \lambda\log(N)$,
where $\lambda$ varies from $0.5$ to $1$. 
The density of the between-block edges is controlled by  $\lambda$: larger $\lambda$ lead to sparsity of between-blocks edges.
In line with our expectations, Figure \ref{fig:phi_combined} shows that the block recovery worsens as the number of between-block edges increases (i.e., when $\lambda$ decreases). 

To ensure that our main results are not sensitive to specific modeling choices or parameter settings, we conducted a series of additional analyses. These include evaluating alternative clustering approaches and  testing the sensitivity of block recovery to the choice of the number of blocks $K$. 

\subsection{Comparison of Alternative Clustering Variants} \label{sec:comp_clust}
In addition to the comparison between variational approximation and binary spectral clustering in Section \ref{sec:sim}, we evaluated a binary spectral clustering variant where negative and zero edges are merged and treated as absent ties. This approach implicitly assumes a fixed relationship between edge sign and block structure, namely that negative edges occur primarily between blocks. While reasonable in some settings, including the one presented in Section \ref{sec:sim}, this constraint limits flexibility in more general signed networks.

\begin{figure}
    \centering
    \includegraphics[width=\linewidth]{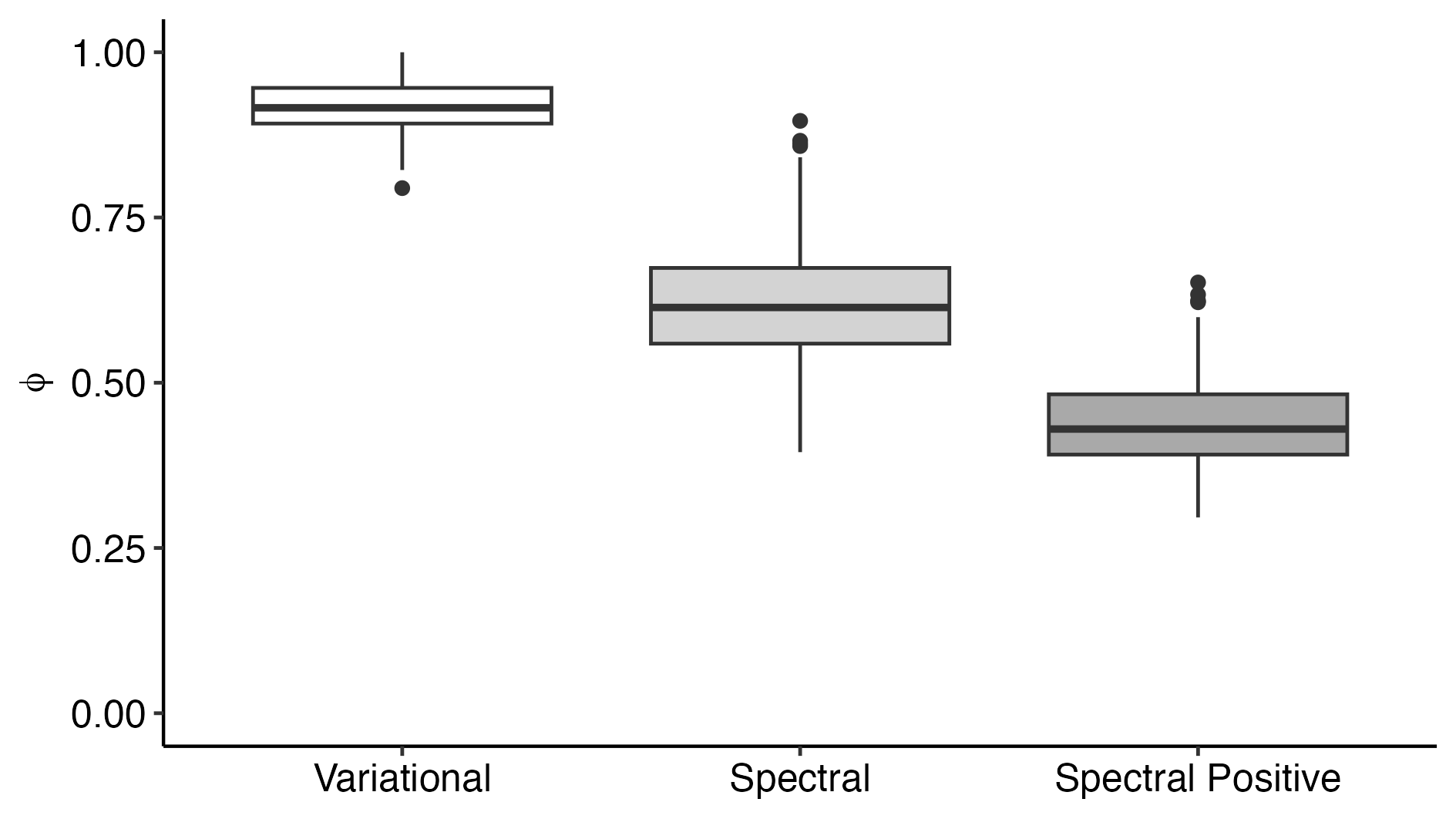}
    \caption{Comparison of variational approximation, binary spectral clustering, and spectral clustering using only positive edges for block recovery.}
    \label{fig:dep_between}
\end{figure}

Merging negative and zero edges substantially increases sparsity, in particular for between-block connections (see Figure~\ref{fig:dep_between}). The resulting loss of information degrades the performance of binary spectral clustering, leading to lower block recovery accuracy than the baseline binary method that ignores edge signs entirely. 

\subsection{Sensitivity to Misspecification of $K$} \label{sec:missk}
\begin{figure}
    \centering
    \includegraphics[width=\linewidth]{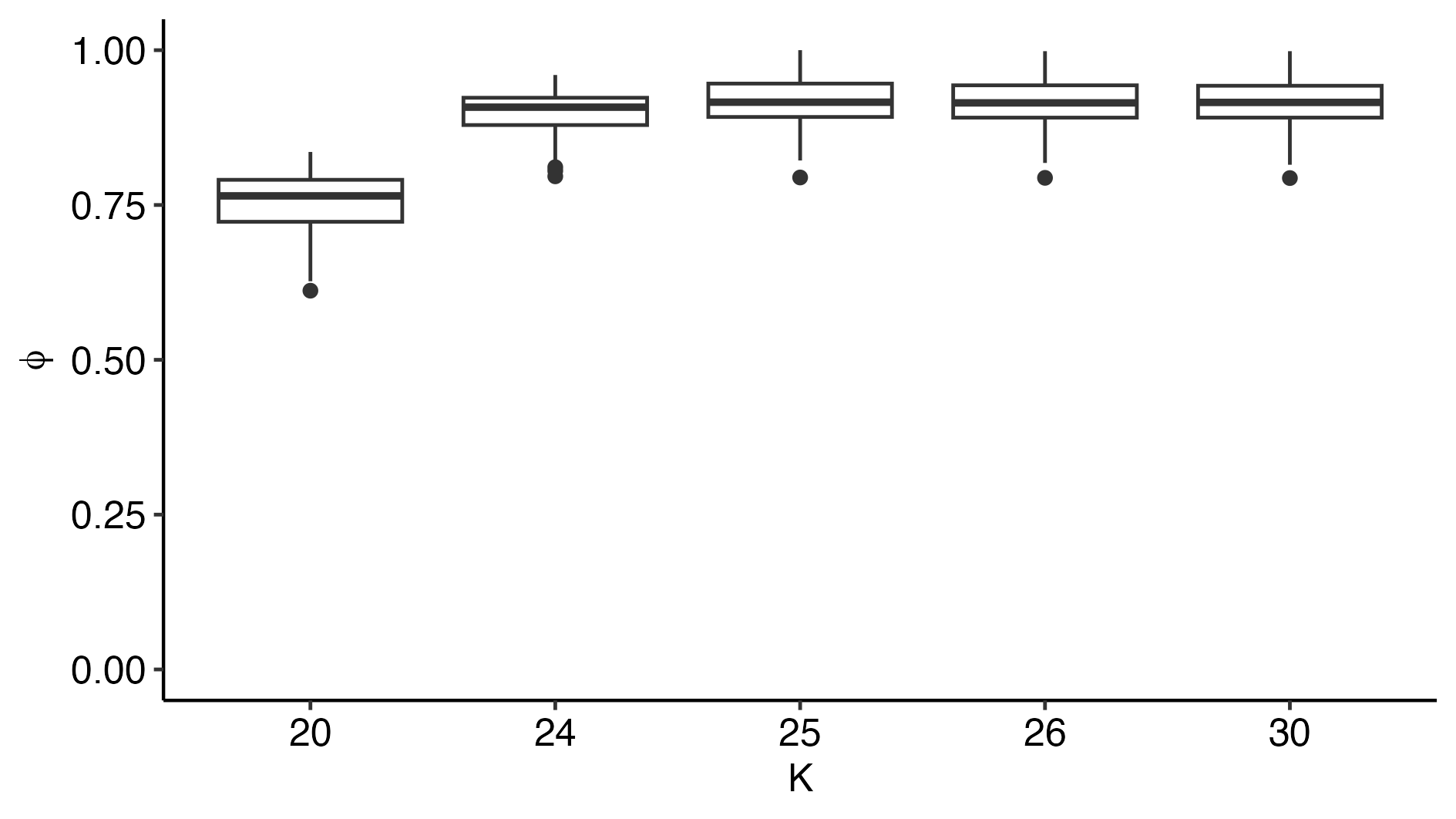}
    \caption{Robustness of block recovery to misspecification of $K=25$. We varied the number of blocks around the true value ($K \pm 1$ and $K \pm 5$) to assess sensitivity.}
    \label{fig:k_sensitivity}
\end{figure}

To assess robustness to misspecification of the number of blocks, we varied $K$ around the true value of $25$, considering both small ($K \pm 1$) and moderate ($K \pm 5$) deviations (see Figure~\ref{fig:k_sensitivity}). The results show a clear asymmetry. When $K$ is overestimated, performance remains essentially unchanged: the additional blocks contain very few nodes and do not affect block recovery or agreement measures. In contrast, underestimating $K$ leads to a gradual decline in performance, as distinct blocks are forced to merge, which necessarily reduces recovery accuracy. Overall, the method is robust to moderate misspecification of $K$, particularly in the direction of overestimation.

\section{Wikipedia Network} \label{sec:app_wiki}
This appendix provides supplementary material for the Wikipedia network application described in Section \ref{sec:irl}. It includes the list of the 50 selected Wikipedia pages, out-of-sample cross-validation as well as in-sample goodness of fit plots and a visualization of the full Wikipedia network. These materials support the evaluation of model fit and illustrate the network structure underlying the analysis.
\begin{table}[!t]
\caption{Network statistics for the Wikipedia editor network, including the number of nodes, positive and negative edges, and frequencies of triads: all-positive (\textsl{``friend of my friend is my friend"}), all-negative, two positive and one negative, and one positive and two negative (\textsl{``enemy of my enemy is my friend"}).}
\begin{center}
\begin{tabular}{lccccccc}
\toprule
Nodes & Edges $+$ &Edges $-$ &
\triadppp &
\triadmmm &
\triadppm &
\triadpmm \\
\midrule
\mbox{2,115}  & 875 & \mbox{2,656} & 78 & 738 & 272 & 934 \\
\bottomrule
\end{tabular}
\end{center}
\end{table}

\begin{table}[!tbp]
\caption{List of selected Wikipedia pages used to construct the editor network.\label{tbl:wiki_pages}}
\vspace{-1pt}
\centering
\scriptsize
\renewcommand{\arraystretch}{0.9}
\begin{tabularx}{\textwidth}{X X}
\hline
\multicolumn{2}{l}{Wikipedia pages} \\
\hline
Second Taranaki War & Operation Nifty Package \\
Te Kooti's War & Box Hill Hawks Football Club \\
2011 Turkish sports corruption scandal & Comparison of MUTCD-influenced traffic signs \\
Turkish Cup & Calogero Vizzini \\
Navenby & Partial pressure \\
Sunderland Echo & Edmund Lyons, 1st Baron Lyons \\
2009-10 Ukrainian First League & John Jervis, 1st Earl of St Vincent \\
2010 S.League & Falcon's Fury \\
King's Cup & Hogwarts Express (Universal Orlando Resort) \\
TT Pro League & Francization of Brussels \\
Akhtar Hameed Khan & Leonel Brizola \\
Ishaq Dar & Glenda Farrell \\
Alias (season 5) & List of Maverick episodes \\
List of The Listener episodes & Herne Hill railway station \\
Argentina women's national field hockey team & LSWR N15 class \\
Australia national baseball team & Hillsborough Area Regional Transit \\
Lena Park & List of state highways in Arkansas \\
Battle of Flamborough Head & The Verge \\
Uzalo & Iveta Mukuchyan \\
Margaret (singer) & Japanese aircraft carrier Hiryū \\
Principalía & Kulothunga Chola III \\
Northern Province, Sri Lanka & Lee Purcell \\
New Guinea singing dog & Page Two (EP) \\
Talbot Tagora & Tatiana Troyanos \\
Little Thetford & Temple of Eshmun \\
\hline
\end{tabularx}
\end{table}

\clearpage
\subsection{Out-of-Sample Cross Validation} \label{sec:oos_sup}

\begin{figure}[!h]
    \centering
     \includegraphics[width=\linewidth]{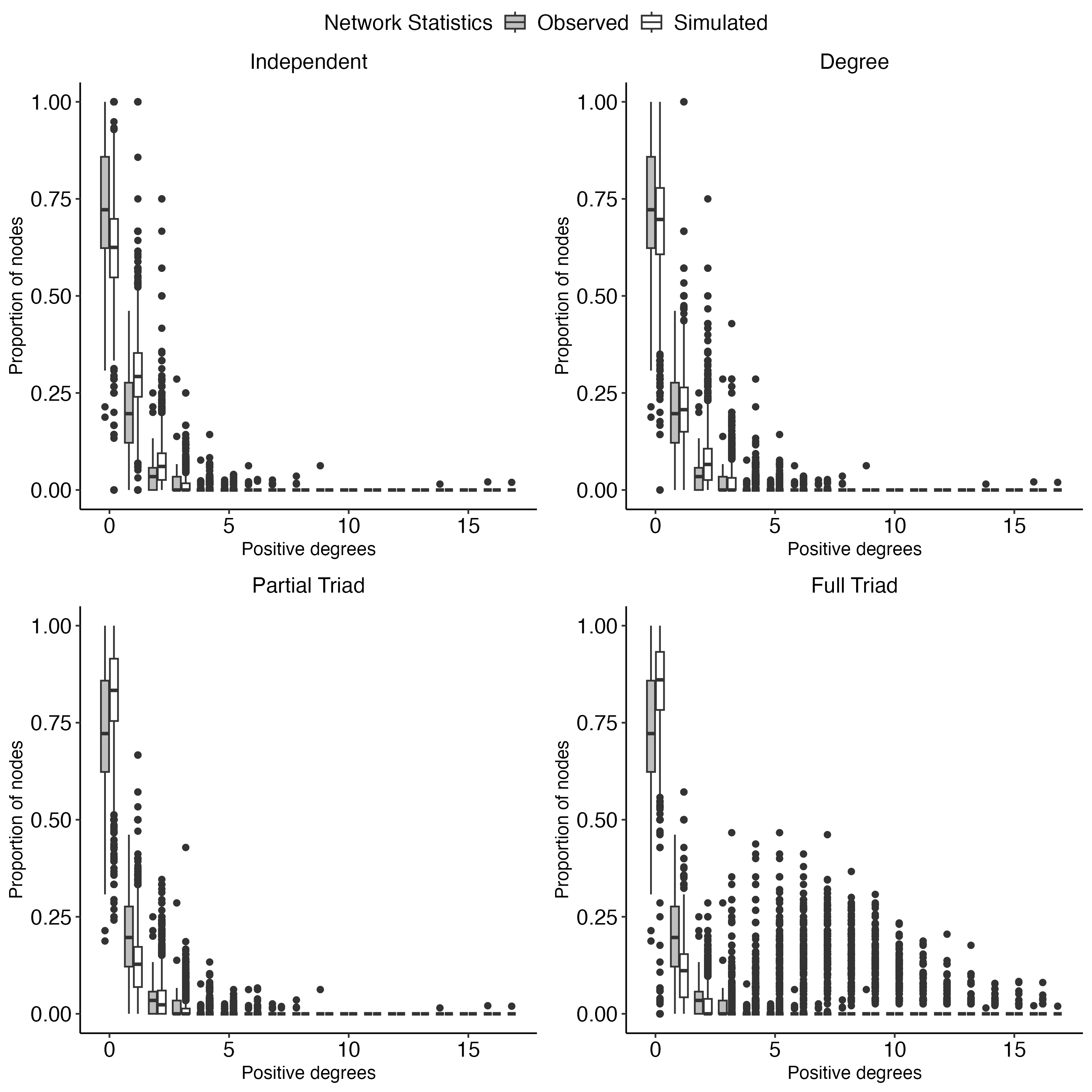}
         \caption{Comparison of out-of-sample cross-validation results for positive degree distribution. The distribution of simulated statistics across 100 replications is compared against the observed statistics for each block. Models compared are: I (Independent), I+D (Degree), I+D+PT (Partial Triad), and I+D+FT (Full Triad).}
    \label{fig:oos_pd}
\end{figure}

\begin{figure}[!h]
    \centering
     \includegraphics[width=\linewidth]{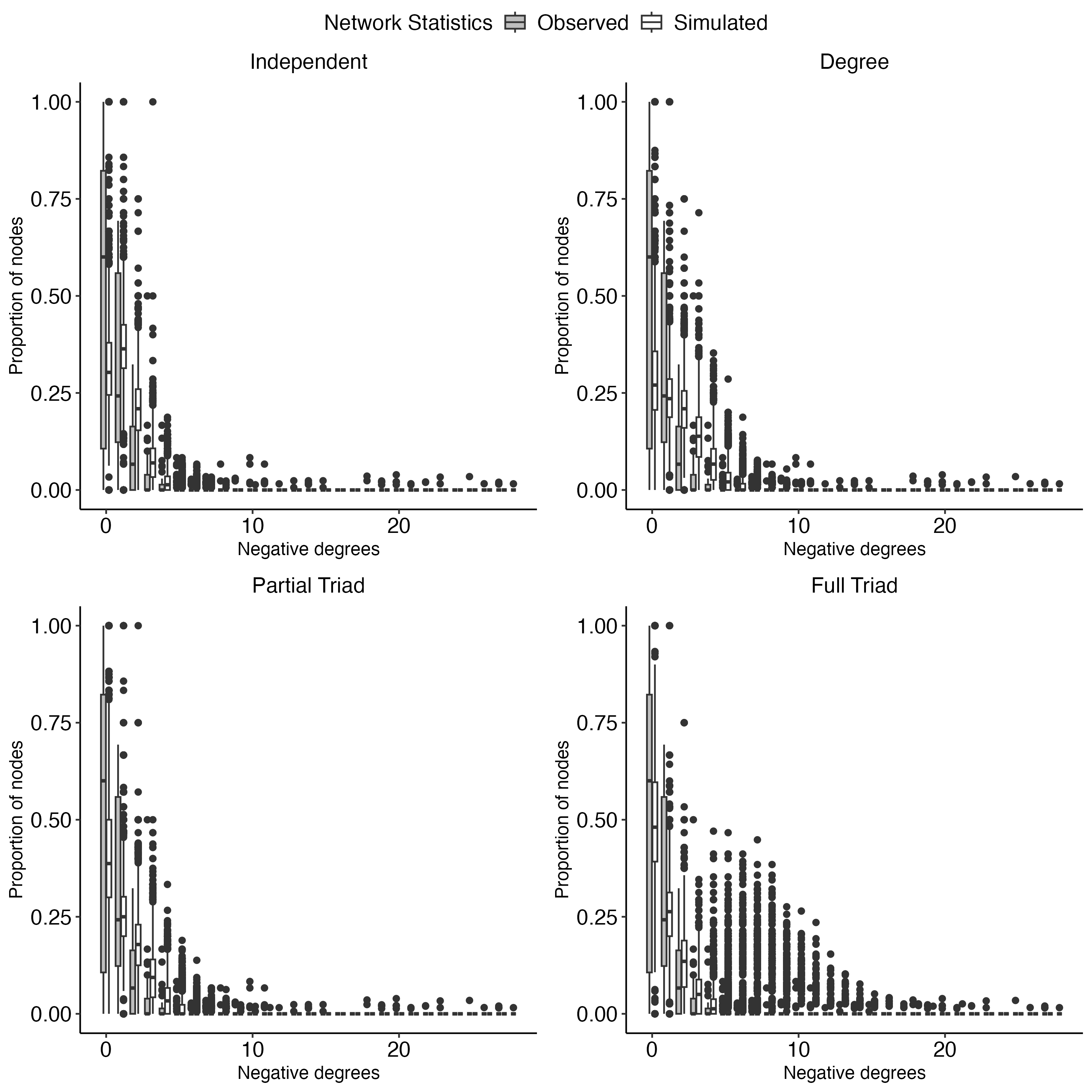}
         \caption{Comparison of out-of-sample cross-validation results for negative degree distribution. The distribution of simulated statistics across 100 replications is compared against the observed statistics for each block. Models compared are: I (Independent), I+D (Degree), I+D+PT (Partial Triad), and I+D+FT (Full Triad).}
    \label{fig:oos_nd}
\end{figure}

\begin{figure}[!h]
    \centering
     \includegraphics[width=\linewidth]{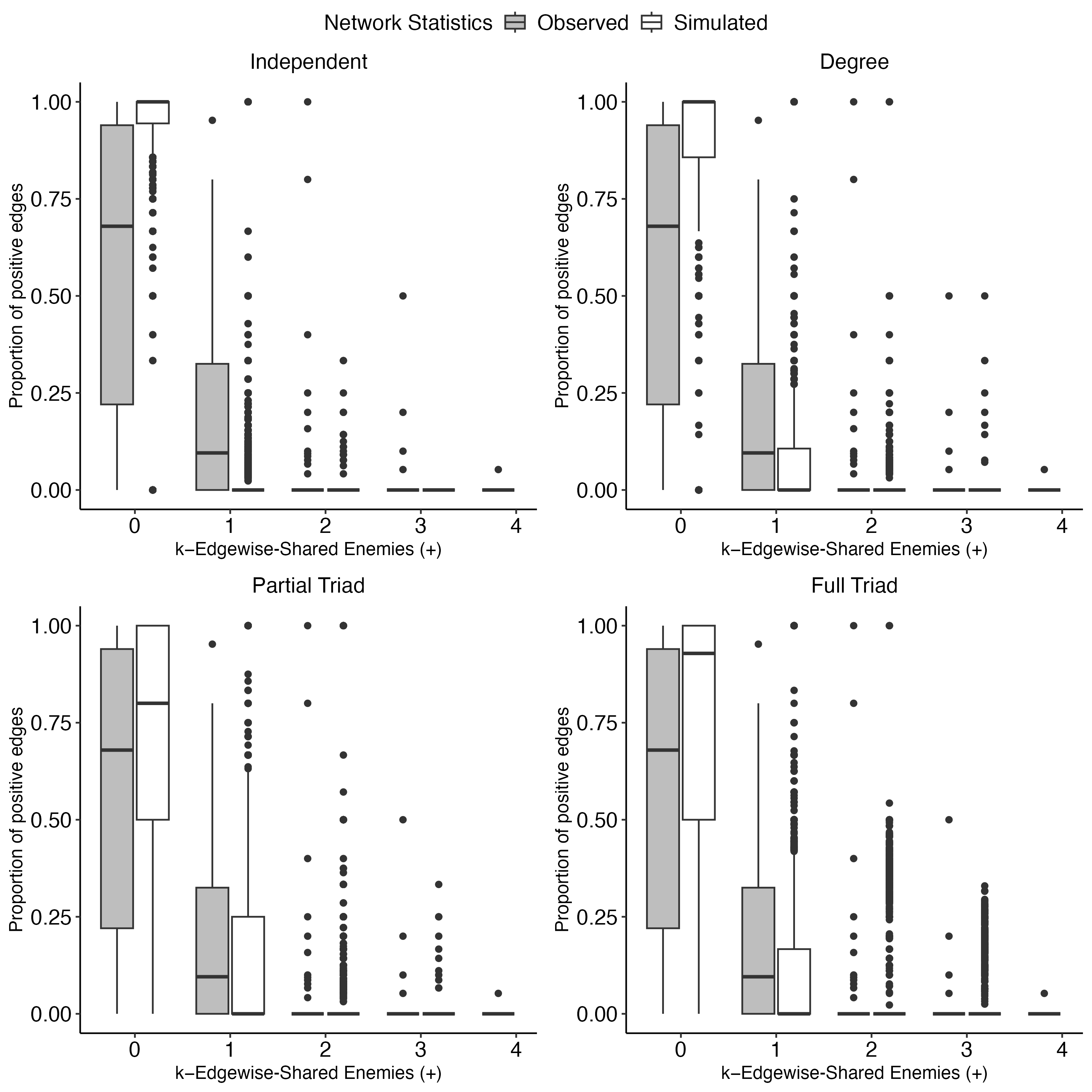}
             \caption{Comparison of out-of-sample cross-validation results for positive edgewise shared enemies (ESE $+$). The distribution of simulated statistics across 100 replications is compared against the observed statistics for each block. Models compared are: I (Independent), I+D (Degree), I+D+PT (Partial Triad), and I+D+FT (Full Triad).}
    \label{fig:oos_gwese_pos}
\end{figure}

\begin{figure}[!h]
    \centering
     \includegraphics[width=\linewidth]{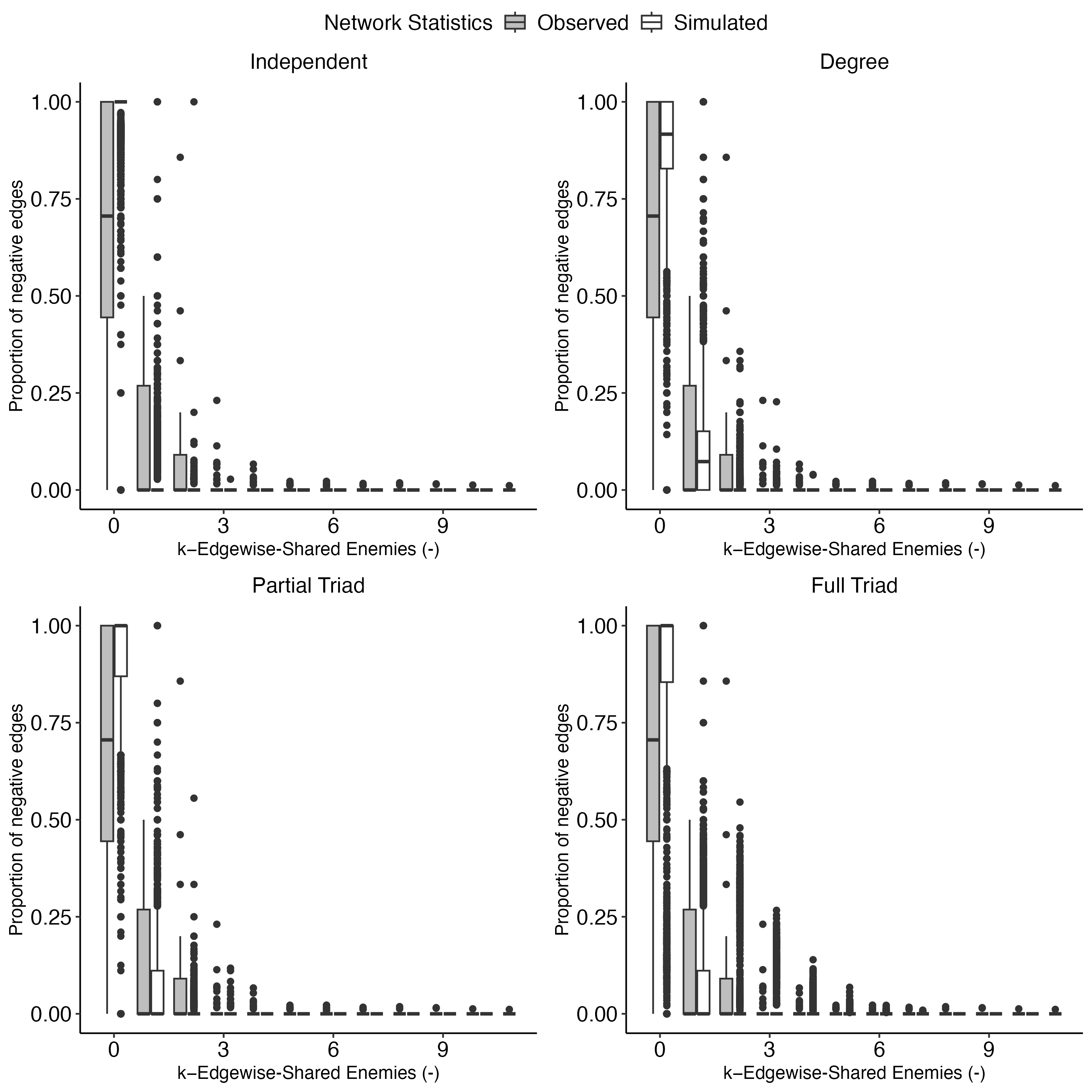}
             \caption{Comparison of out-of-sample cross-validation results for negative edgewise shared enemies (ESE $-$). The distribution of simulated statistics across 100 replications is compared against the observed statistics for each block. Models compared are: I (Independent), I+D (Degree), I+D+PT (Partial Triad), and I+D+FT (Full Triad).}
    \label{fig:oos_gwese_neg}
\end{figure}

\begin{figure}[!h]
    \centering
     \includegraphics[width=\linewidth]{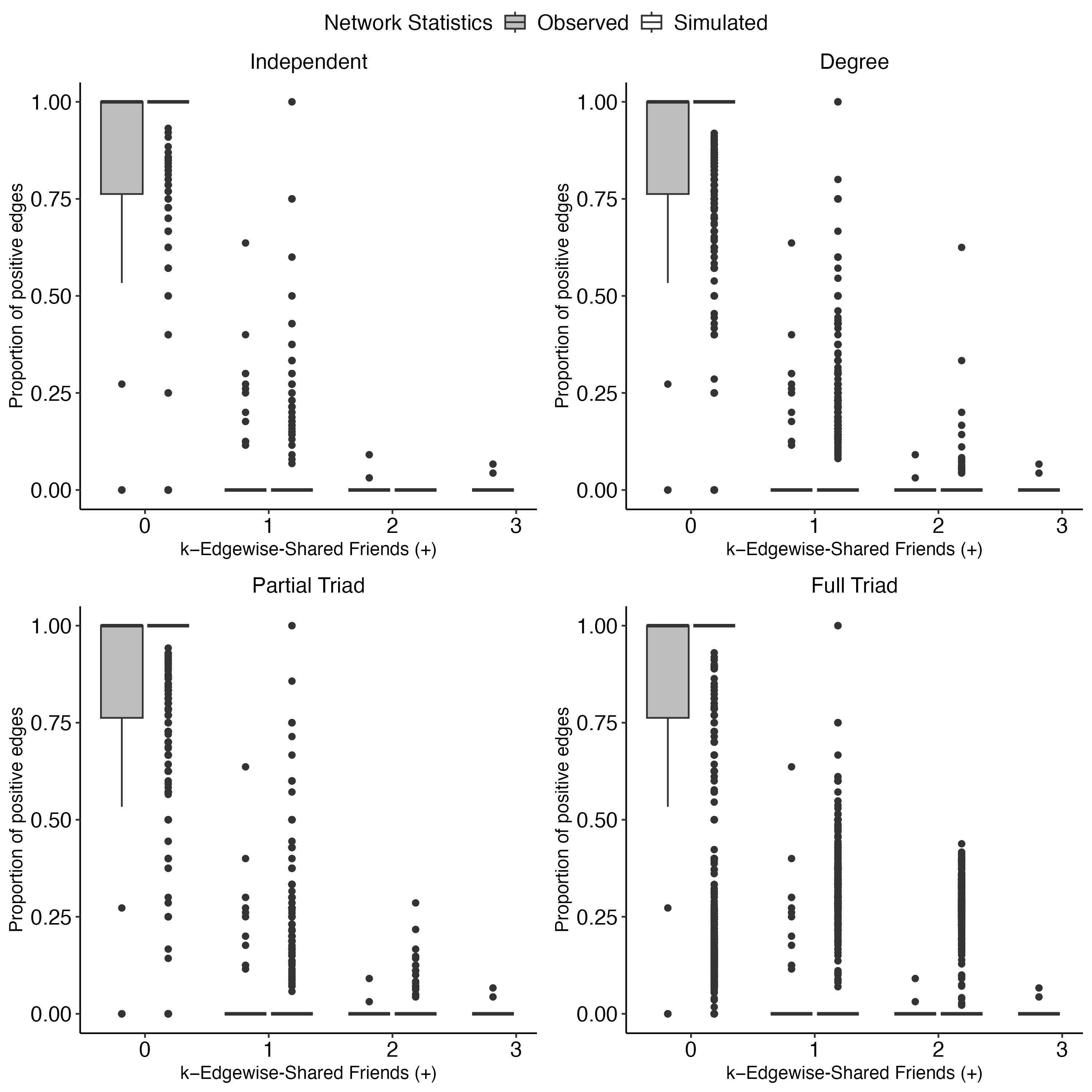}
         \caption{Comparison of out-of-sample cross-validation results for positive edgewise shared friends (ESF $+$). The distribution of simulated statistics across 100 replications is compared against the observed statistics for each block. Models compared are: I (Independent), I+D (Degree), I+D+PT (Partial Triad), and I+D+FT (Full Triad).}
    \label{fig:oos_gwesf_pos}
\end{figure}

\begin{figure}[!h]
    \centering
     \includegraphics[width=\linewidth]{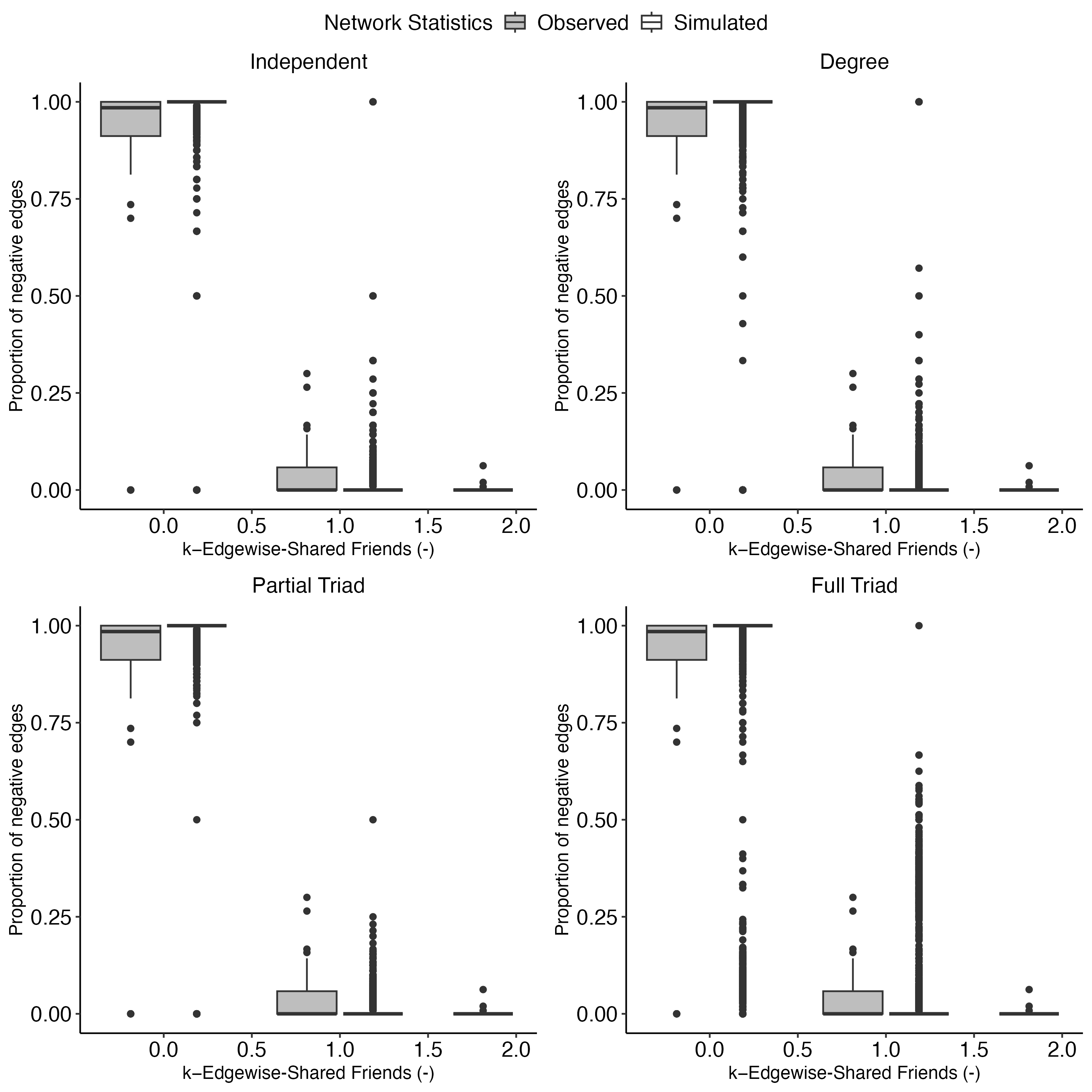}
     \caption{Comparison of out-of-sample cross-validation results for negative edgewise shared friends (ESF $-$). The distribution of simulated statistics across 100 replications is compared against the observed statistics for each block. Models compared are: I (Independent), I+D (Degree), I+D+PT (Partial Triad), and I+D+FT (Full Triad).}
    \label{fig:oos_gwesf_neg}
\end{figure}

\clearpage

\subsection{Goodness-of-Fit} \label{sec:gof_sup}
To validate the results in Table \ref{tbl:ergm_within_between}, we conduct a conventional ERGM goodness-of-fit analysis following the method outlined by \citesupp{hunter2008goodness}. 
Using the estimated coefficients, we simulate 500 networks and compare the simulated network statistics with the observed statistics. For the network statistics describing the signed network, we use the same metrics as in Section \ref{sec:oos}. The red line marks the observed network's statistic and should ideally be near the median of the simulated values, shown by the center of the boxplots.

The results show that while the sparse, low-degree nature of the Wikipedia network is captured by the simulations, the statistics distribution of the observed network is not replicated with high accuracy.
Overall, the in-sample goodness-of-fit analysis aligns with the out-of-sample analysis, indicating that the more complex models, ``Partial Triad" and ``Full Triad", best describe the data.

\begin{figure} [!h]
    \centering
     \includegraphics[width=\linewidth]{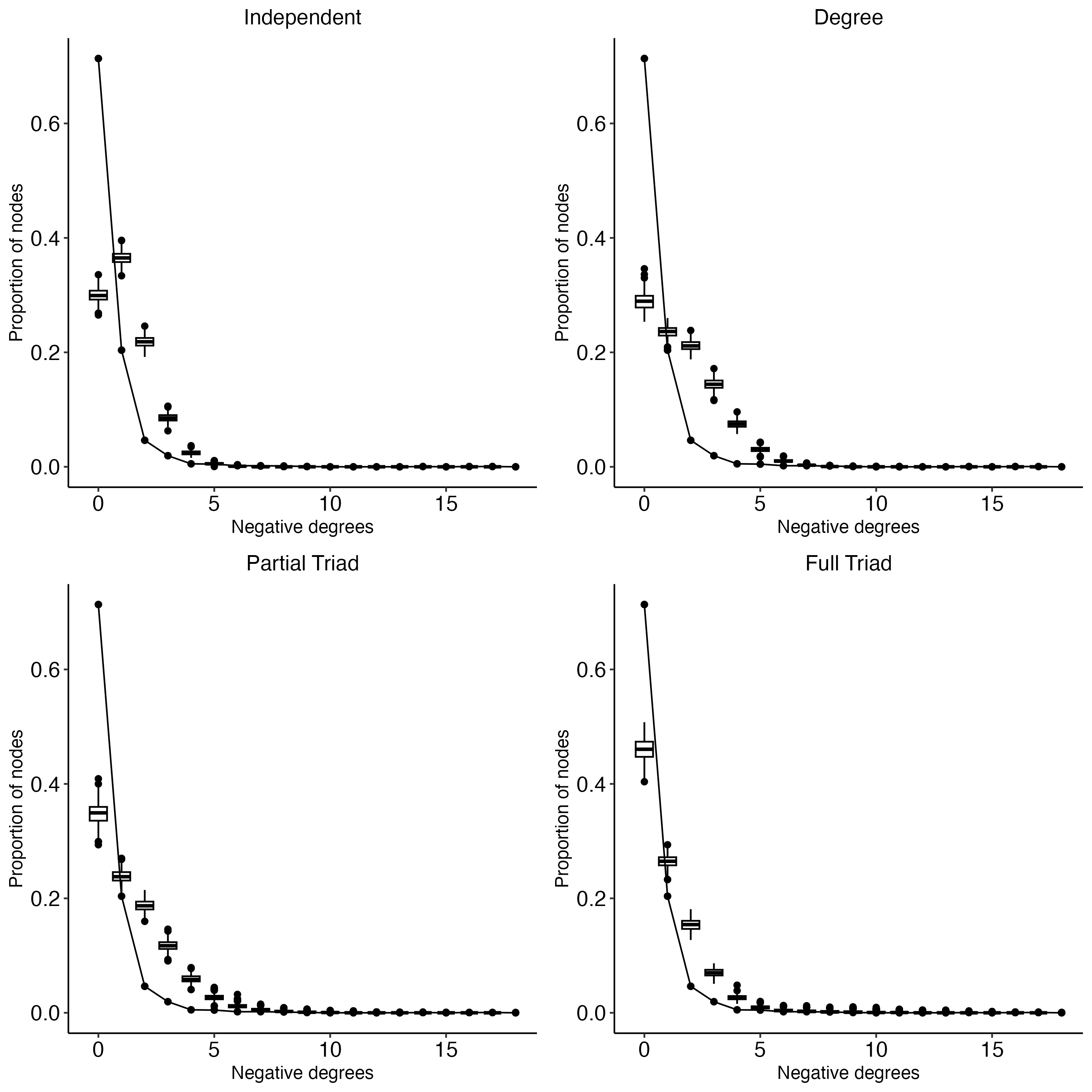}
         \caption{Comparison of goodness-of-fit results for positive degree distribution. Observed network statistics (line) are compared to the distribution of statistics from 500 simulated networks. Models compared are: I (Independent), I+D (Degree), I+D+PT (Partial Triad), and I+D+FT (Full Triad).}
    \label{fig:gof_degree_pos}
\end{figure}

\begin{figure}[!h]
    \centering
     \includegraphics[width=\linewidth]{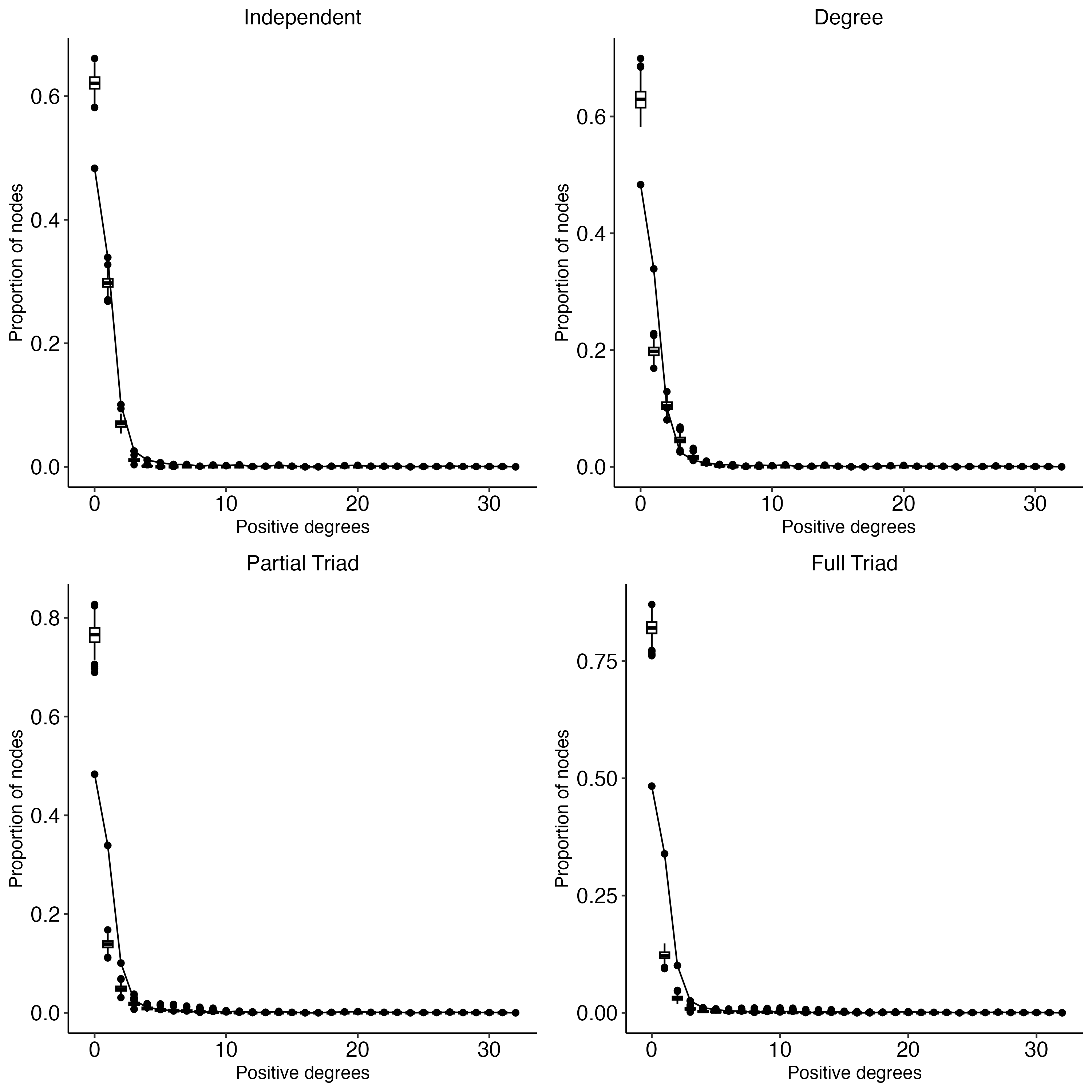}
             \caption{Comparison of goodness-of-fit results for negative degree distribution. Observed network statistics are compared to the distribution of statistics from 500 simulated networks. Models compared are: I (Independent), I+D (Degree), I+D+PT (Partial Triad), and I+D+FT (Full Triad).}
    \label{fig:gof_degree_neg}
\end{figure}

\begin{figure}[!h]
    \centering
     \includegraphics[width=\linewidth]{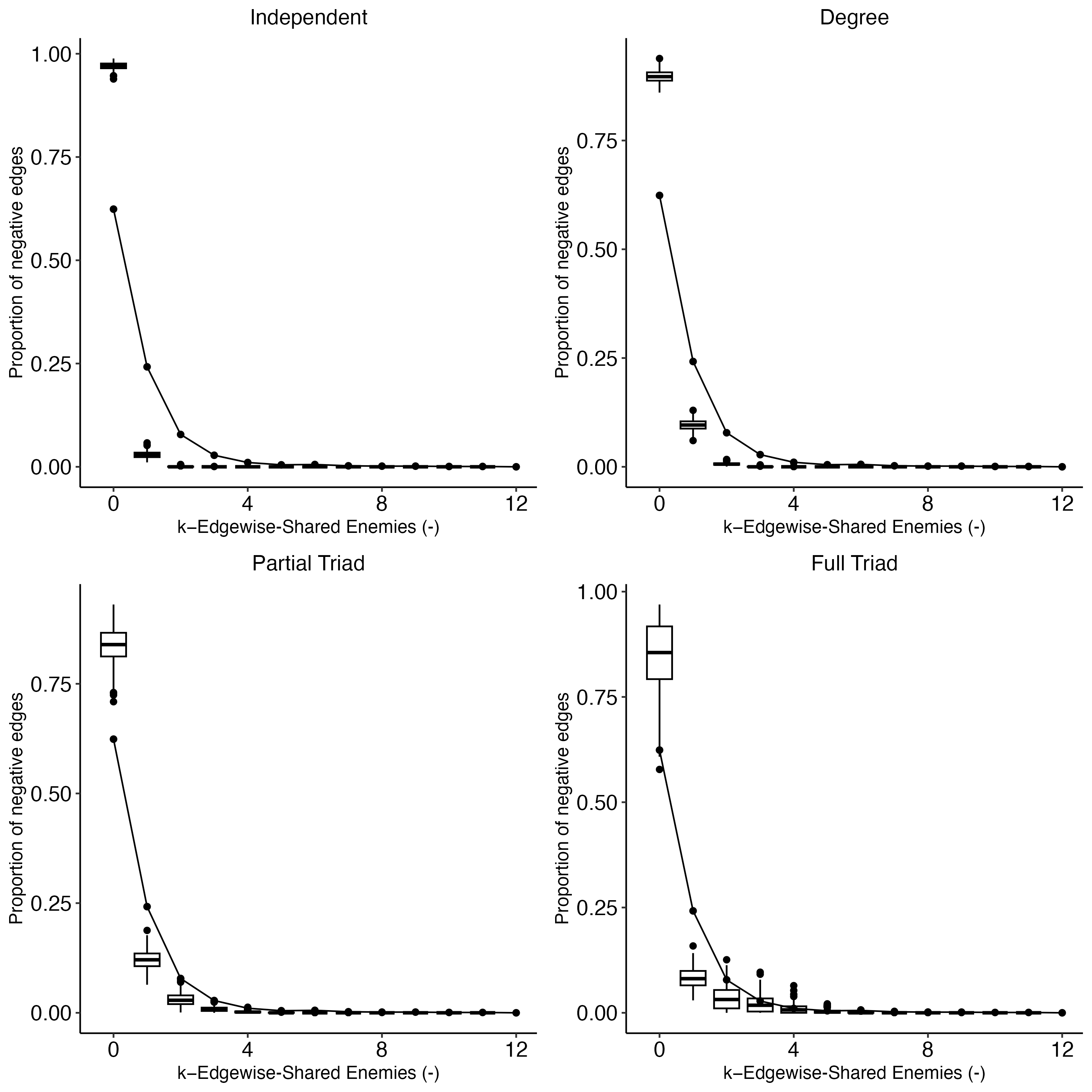}
             \caption{Comparison of goodness-of-fit results for positive edgewise shared enemies (ESE $+$). Observed network statistics are compared to the distribution of statistics from 500 simulated networks. Models compared are: I (Independent), I+D (Degree), I+D+PT (Partial Triad), and I+D+FT (Full Triad).}
    \label{fig:gof_ese_pos}
\end{figure}

\begin{figure}[!h]
    \centering
     \includegraphics[width=\linewidth]{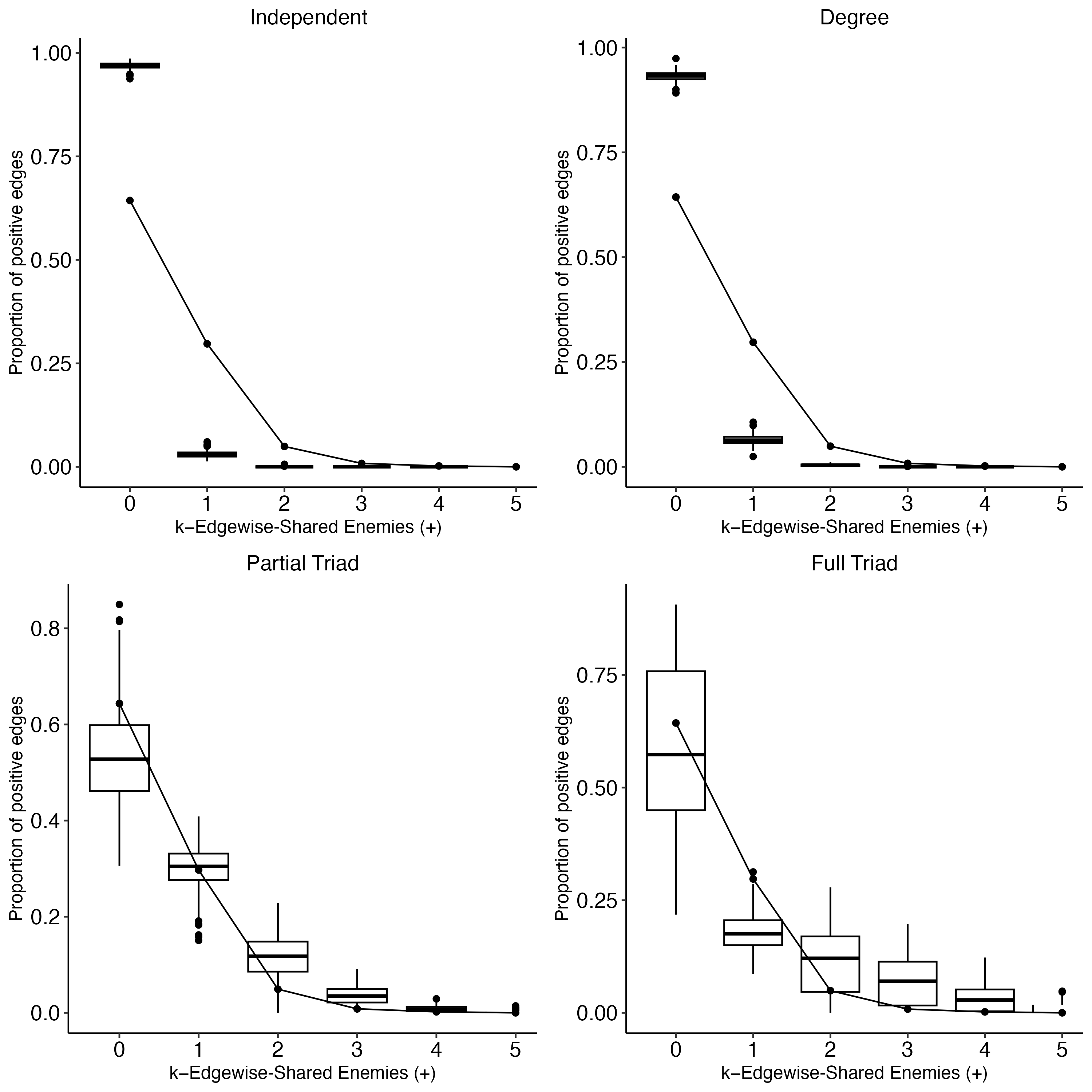}
             \caption{Comparison of goodness-of-fit results for negative edgewise shared enemies (ESE $-$). Observed network statistics are compared to the distribution of statistics from 500 simulated networks. Models compared are: I (Independent), I+D (Degree), I+D+PT (Partial Triad), and I+D+FT (Full Triad).}
    \label{fig:gof_ese_neg}
\end{figure}

\begin{figure}[!h]
    \centering
     \includegraphics[width=\linewidth]{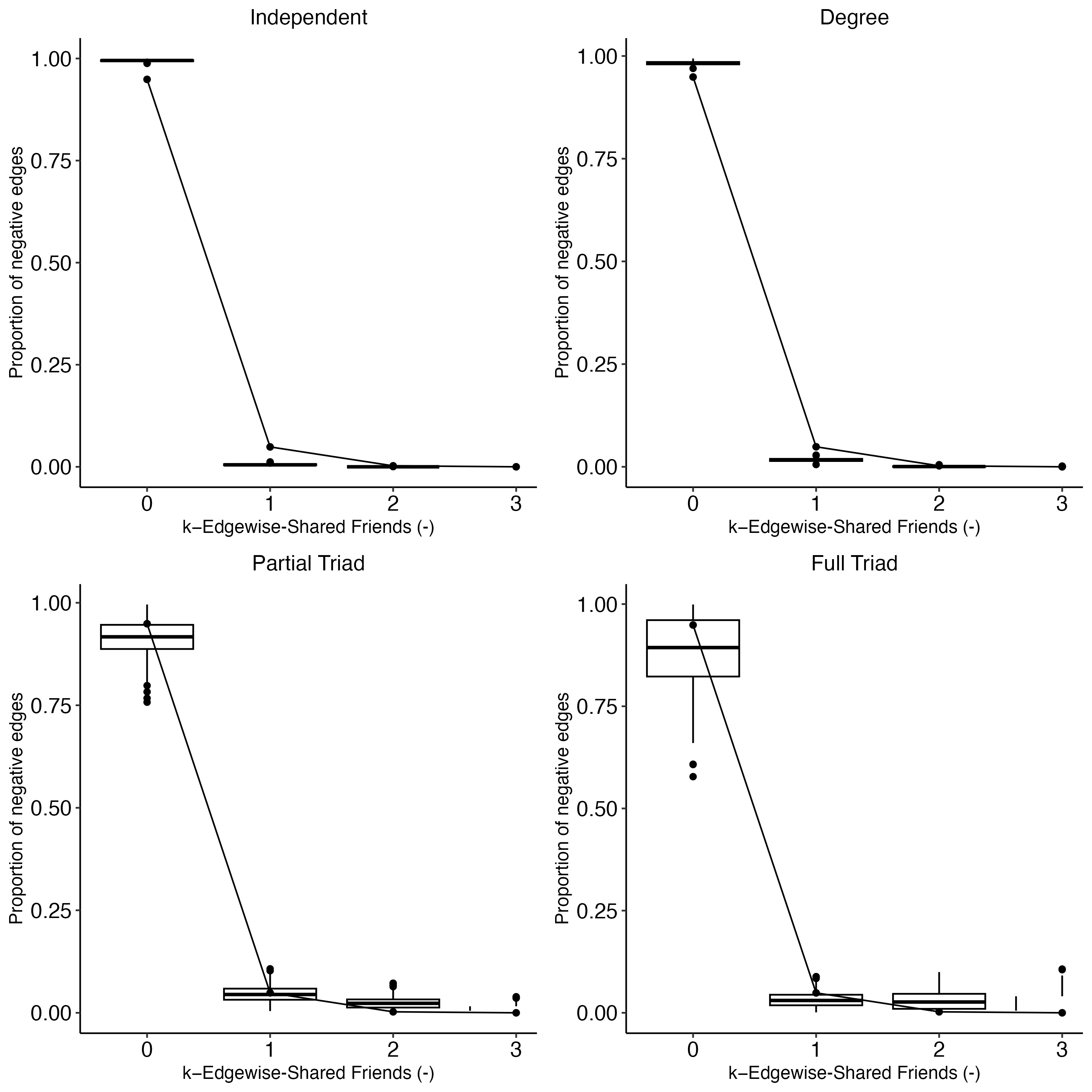}
             \caption{Comparison of goodness-of-fit results for positive edgewise shared friends (ESF $+$). Observed network statistics are compared to the distribution of statistics from 500 simulated networks. Models compared are: I (Independent), I+D (Degree), I+D+PT (Partial Triad), and I+D+FT (Full Triad).}
    \label{fig:gof_esf_pos}
\end{figure}

\begin{figure}[!h]
    \centering
     \includegraphics[width=\linewidth]{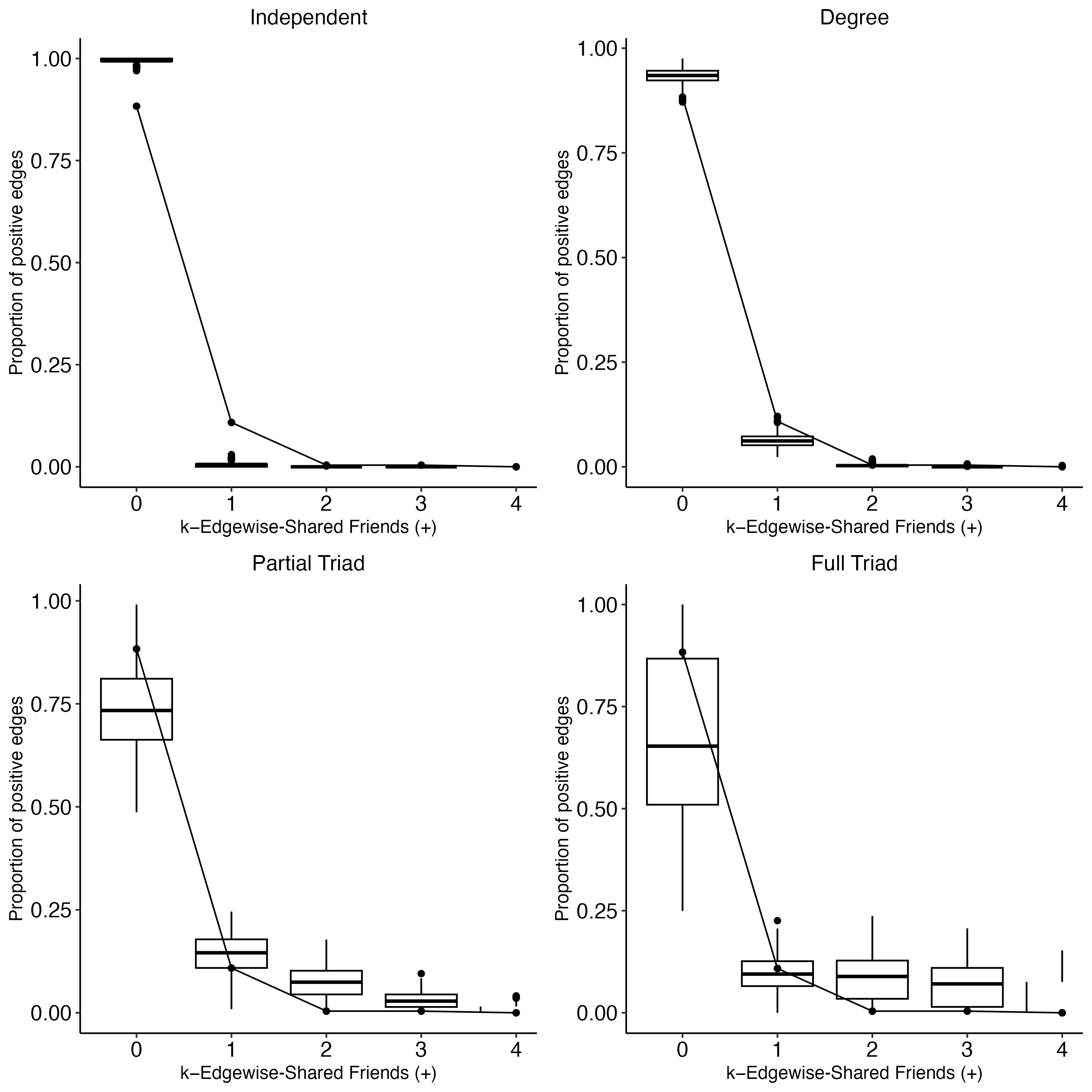}
             \caption{Comparison of goodness-of-fit results for negative edgewise shared friends (ESF $-$). Observed network statistics are compared to the distribution of statistics from 500 simulated networks. Models compared are: I (Independent), I+D (Degree), I+D+PT (Partial Triad), and I+D+FT (Full Triad).}
    \label{fig:gof_esf_neg}
\end{figure}

\begin{figure}[!h]
    \centering
    \rotatebox{90}{%
        \includegraphics[width=0.7\linewidth]{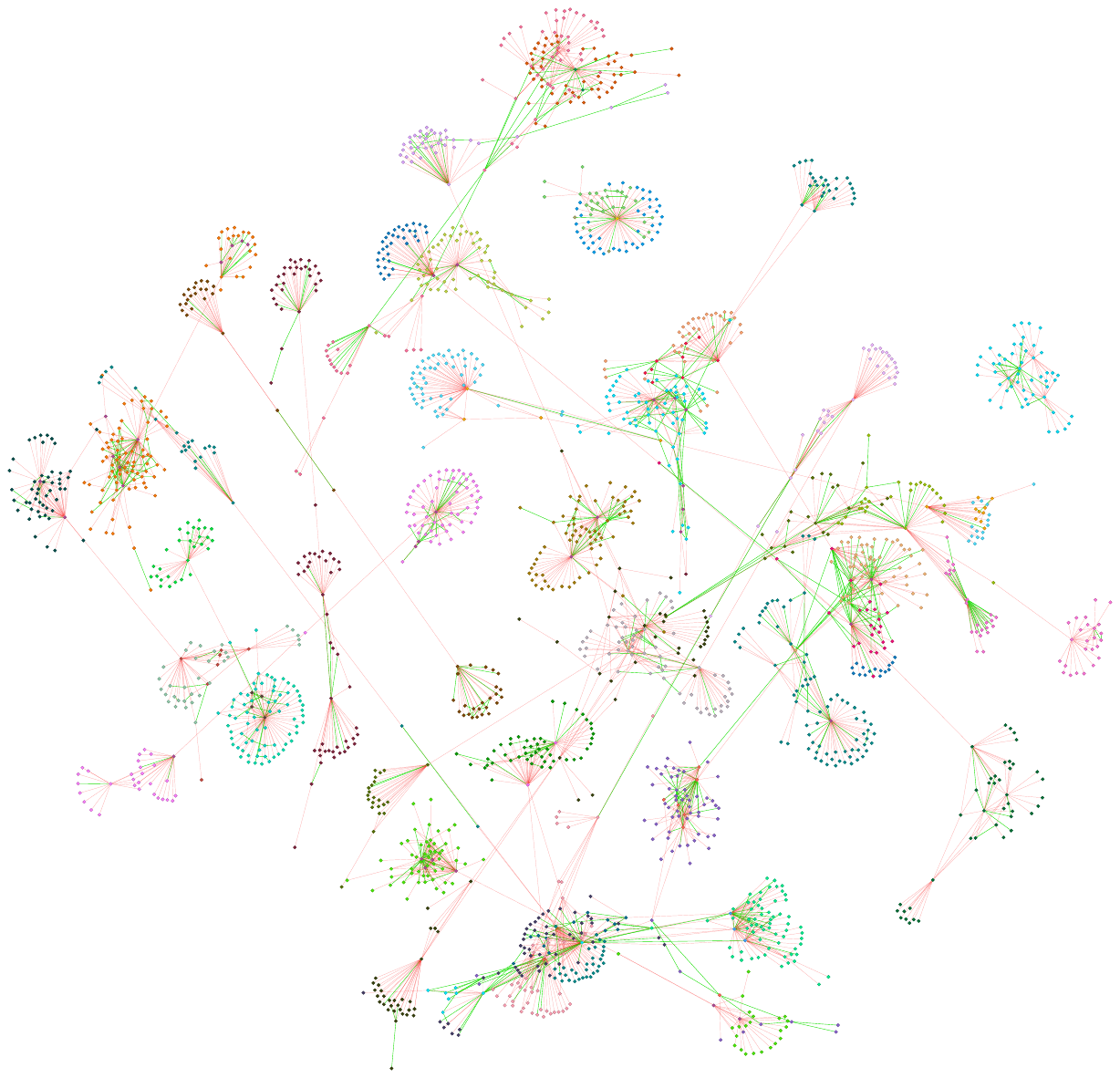}
    }
    \label{fig:enter-label}
        \caption{Visualization of the Wikipedia editor network based on selected pages.}

\end{figure}

\newpage
\clearpage

\subsection{Comparison of Dependence Assumptions} \label{sec:comp_dep}

We evaluate three models to assess how different assumptions about dependence affect the resulting network structure, using the Wikipedia network introduced in Section~\ref{sec:app_wiki}. The estimated coefficients are reported in Table~\ref{tab:model_coefficients}. First, we estimate a standard ERGM across all edges that ignores block structure and assumes global dependence across all dyads (Global Dependence). Second, we consider a model that accounts for block structure by distinguishing between within- and between-block edges and fitting two separate models (Between Dependence). In contrast to our proposed model, this model also characterizes between-block edges using an ERGM rather than an SBM. We compare both alternatives to the fully specified model described in Section~\ref{sec:app_wiki}.

As expected, the Global Dependence model implies a sparse network when averaging over all dyads, with strongly negative coefficients for both positive and negative edges. In the Between Dependence model, the between-block edge coefficients are also strongly negative, but less so than in the Full Triad model, since edge formation can be partly explained by other dyad-dependent terms. Although the coefficients of the triadic terms differ in magnitude, they all convey a similar picture.

These estimated coefficients were used to simulate 500 networks in order to evaluate goodness of fit by comparing observed network statistics to simulated ones. The results are reported in Table~\ref{tab:model_comparison}. We did not use the traditional goodness-of-fit test described in Section~\ref{sec:gof_sup} because edge-normalized statistics make it difficult to compare model fit in sparse networks such as the Wikipedia network. The Global Dependence model strongly underestimates edge and triad counts, whereas the Between Dependence model substantially overestimates both edges and triads. In contrast, the Full Triad model yields simulated statistics that are consistently closer to the observed values. 

These findings suggest that while between-block triads exist and may relate to structural balance theory, explicitly modeling them through higher-order dyad-dependent terms introduces more bias than signal in large-scale networks. 
For the Wikipedia network, excluding cross-block dependencies avoids severe misspecification and appears sufficient to capture the essential structural features.
We note that this result may be network-specific. Extensions incorporating selective or constrained cross-block dependencies remain a promising avenue for future research.

\begin{table}[htbp]
\centering
\caption{Model coefficients used in the simulations. The Global Dependence model specifies a single ERGM for all edges. The Between Dependence model specify separate ERGMs for within- and between-block edges. The Full Triad is the Local Dependence model that proved to be the best fit for the Wikipedia network.}
\label{tab:model_coefficients}
\begin{tabular}{@{}lccccc@{}}
Term 
& Global Dep.
& \multicolumn{2}{c}{Between Dep.}
& \multicolumn{2}{c}{Full Triad} \\
\cmidrule(l){3-4} \cmidrule(l){5-6}  & 
& Within & Between
& Within & Between \\
\hline
$\text{Edges}^+$  & -7.90 &0.3&-5.74& 0.3 & -8.63 \\
\quad$\times\log(N_k)$ & & -0.82& & -0.82 \\
$\text{Edges}^-$  & -8.01 &0.69 & -6.53 & 0.69 & -7.34\\
\quad$\times\log(N_k)$ & & -0.86 & & -0.86 \\
$\text{GWD}^+$   & -0.89& -1.36 & -2.3 & -1.36 \\
$\text{GWD}^-$   & 1.35 & -0.82 & -1.03 & -0.82 \\
 $\text{GWESE}^+$ & 4.12 & 1.60 & 3.85 & 1.60 \\
$\text{GWESF}^+$  & 0.50 & 0.10 & 0.67 & 0.10 \\
$\text{GWESE}^-$  & 1.28 & 0.19 & 0.72 & 0.19 \\
$\text{GWESF}^-$  & 1.72 & 0.81 & 1.60 & 0.81 \\
\hline
\end{tabular}
\end{table}

\begin{table}[htbp]
\centering
\caption{Comparison of observed vs. simulated network statistics. Mean and standard deviation (SD) are calculated across 500 simulated networks for each model. 
Global Dep. represents a global ERGM that ignores local dependence and Between Dep. represents a model where the between block edges are also modeled by an ERGM instead of a SBM.}
\label{tab:model_comparison}
\begin{tabular}{@{}lrrrrr@{}}
& Observed & Global Dep. & Between Dep. & Full Triad \\ 
\hline
\quad Edges (total)& 3,531 & 1,833 (34.7) & 77,169 (4,228) & 3,234 (253) \\
\quad Edges+ & 875 & 205 (17) & 42,472 (2,263) & 761 (138)  \\
\quad Edges$-$ & 2,656 & 1628 (29.4) & 34,698 (1,970) & 2,474 (121)  \\
\addlinespace
\quad Triads (total) & 2,022 & 22.6 (4.99)  & 119,890 (14,676) & 587 (549)\\
\quad + + + & 78 & 0.036 (0.19) & 18,925 (2,251) & 126 (127) \\
\quad - - - & 738 & 11.8 (3.64) & 10,987 (1,455) & 112 (88.3)\\
\quad + + - & 272 & 0.3 (0.54) & 43,315 (5,607) & 364 (353) \\
\quad + - - & 934 & 10.4 (3.26) & 48,097 (5,501)& 364 (326)  \\
\hline
\end{tabular}
\end{table}

\clearpage
\bibliographystylesupp{chicago}
\bibliographysupp{references}
\end{document}